%% file: main.tex
\documentclass[9pt,journal]{IEEEtran}
\usepackage[utf8]{inputenc}
\usepackage[english]{babel}
\usepackage{graphicx}
\usepackage{amsmath}
\usepackage{tikz}
\usepackage{array}
\usepackage{booktabs}
\usepackage{float}
\usepackage{amssymb} %\mathcal{}
\usepackage{mathrsfs} %\mathscr{}
\usepackage{xargs}
\usepackage{diagbox}
\usepackage{siunitx} %\SI
\usepackage{bm}
\usepackage{soul}  % use \st{works}, where works will be marked with a medium line across the works
\usepackage{bbm}
\usepackage{tabu}

\usepackage{tikzscale}

\usepackage{mathtools} %for stacking equal sign
\usepackage{commath} %for abs value and norm
\usepackage{hyperref}
\definecolor{darkblue}{RGB}{0, 5, 150}

\usepackage{tikz}
\definecolor{gold}{rgb}{0.85,.66,0}
\definecolor{cian}{rgb}{.02,.7,.95}

\definecolor{orange}{RGB}{210, 105, 30}

\tikzstyle{every plot}=[prefix=plots/pgf-]
\tikzstyle{shape example}=[color=black!30,draw,fill=yellow!30,line width=.5cm,inner xsep=2.5cm,inner ysep=0.5cm]
\usepackage[utf8]{inputenc}

\allowdisplaybreaks

\newcommand{\vectornorm}[1]{\left\lVert#1\right\rVert}
\newcommand{\vectornormsmall}[1]{\lVert#1\rVert}

\newcommand{\vectornormbig}[1]{\big\lVert#1\big\rVert}

\begin{document}
\title{Massive MIMO for Internet of Things (IoT) Connectivity}
\author{\IEEEauthorblockN{Alexandru-Sabin Bana$^1$, Elisabeth de Carvalho$^1$, Beatriz Soret$^1$, \\ Taufik Abr\~ao$^2$, Jos\'e Carlos Marinello$^2$, Erik G. Larsson$^3$, and Petar Popovski$^1$}\\
\IEEEauthorblockA{$^1$Department of Electronic Systems, Aalborg University, Denmark}\\
\IEEEauthorblockA{$^2$Electrical Engineering Department, Londrina State University, Parana, Brazil}\\
\IEEEauthorblockA{$^3$Department of Electrical Engineering (ISY), Link\"oping University, 581 83 Link\"oping, Sweden}
\thanks{$^1$The authors have been in part supported the European Research Council (ERC) under the European Union Horizon 2020 research and innovation program (ERC Consolidator Grant Nr. 648382 WILLOW) and Danish Council for Independent Research (Grant Nr. 8022-00284B SEMIOTIC).} 
\thanks{$^2$The authors have been supported in part by the CONFAP-ERC Agreement H2020 (Brazilian National Council of State Funding Agencies and European Research Council) and in part by the National Council for Scientific and Technological Development (CNPq) of Brazil under Grant 304066/2015-0 and in part by the Coordenaç\~{a}o de Aperfeiçoamento de Pessoal de Nível Superior - Brazil (CAPES) - Finance Code 001.}
\thanks{$^3$The work of Erik G. Larsson has been supported by the Swedish Research Council (VR) and ELLIIT.}
}

\maketitle
% \tableofcontents
\begin{abstract}
Massive MIMO is considered to be one of the key technologies in the emerging 5G systems, but also a concept applicable to other wireless systems. 
Exploiting the large number of degrees of freedom (DoFs) of massive MIMO essential for achieving high spectral efficiency, high data rates and extreme spatial multiplexing of densely distributed users.
On the one hand, the benefits of applying massive MIMO for broadband communication are well known and there has been a large body of research on designing communication schemes to support high rates. On the other hand, using massive MIMO for Internet-of-Things (IoT) is still a developing topic, as IoT connectivity has requirements and constraints that are significantly different from the broadband connections. In this paper we investigate the applicability of massive MIMO to  IoT connectivity. Specifically, we treat the two generic types of IoT connections envisioned in 5G: massive machine-type communication (mMTC) and ultra-reliable low-latency communication (URLLC). 
This paper fills this important gap by identifying the opportunities and challenges in exploiting massive MIMO for IoT connectivity. 
We provide insights into the trade-offs that emerge when massive MIMO is applied to mMTC or URLLC and present a number of suitable communication schemes. The discussion continues to the questions of network slicing of the wireless resources and the use of massive MIMO to simultaneously support IoT connections with very heterogeneous requirements. The main conclusion is that massive MIMO can bring benefits to the scenarios with IoT connectivity, but it requires tight  integration of the physical-layer techniques with the protocol design. 
\\
\end{abstract}

\begin{IEEEkeywords} mMTC; URLLC; Massive MIMO; 5G; Activity detection; Collision resolution; Extended coverage; short packets; NR; Random Access (RA); Grant-based RA; Grant-free RA; Unsourced RA; Network Slicing; compressing sensing; sparsification; covariance methods; Cross-layer optimization design
\end{IEEEkeywords}

\vspace{4mm}

%%%%%%%%%%%%%%%%%%%%%%%%
\section{Introduction}
%%%%%%%%%%%%%%%%%%%%%%%%

\subsection{The Heterogeneous 5G services}

\begin{figure}[t]
    \centering
    \includegraphics[width=1\linewidth]{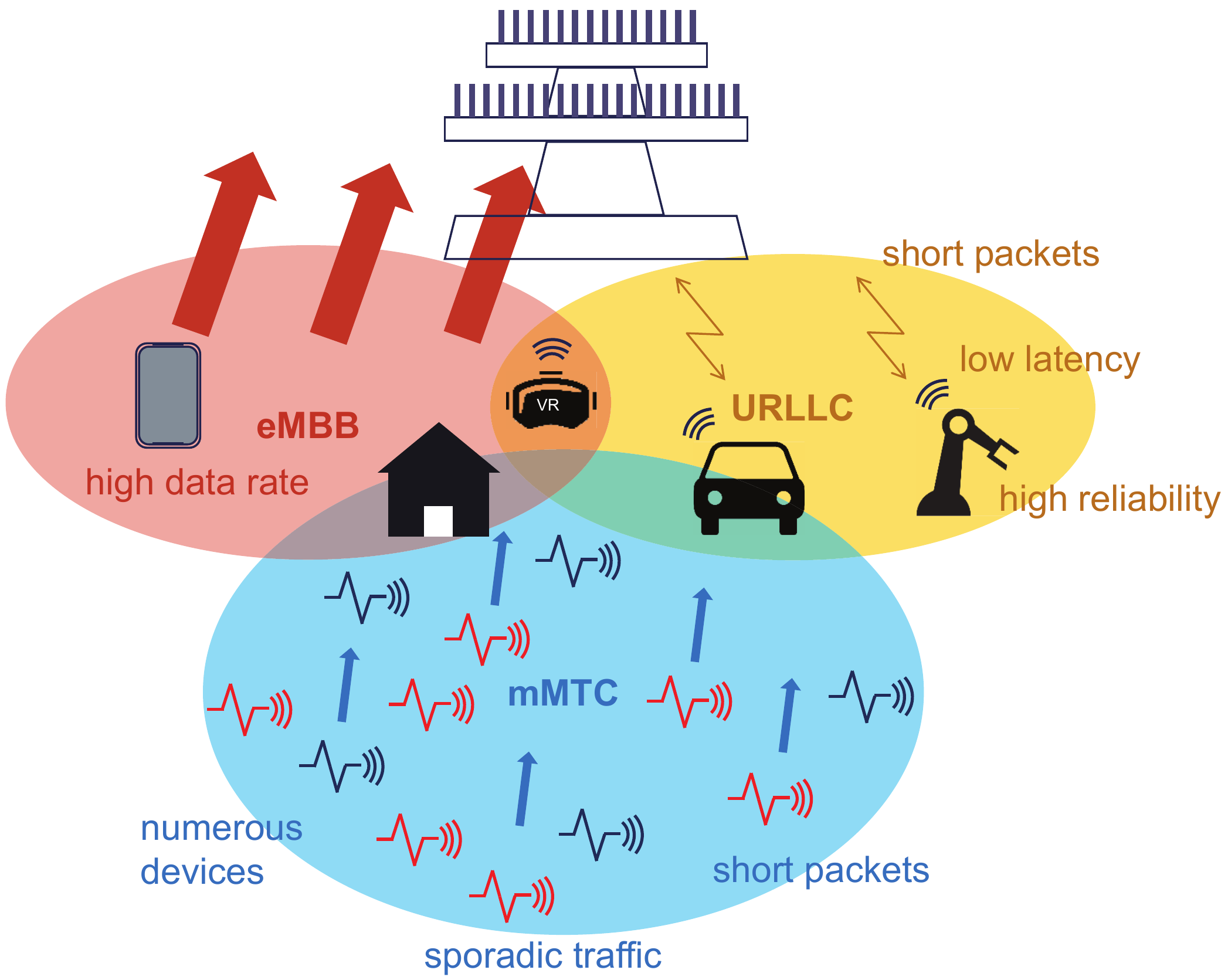}
    \caption{Massive MIMO and the 5G services.}
    \label{fig:scenario3}
\end{figure}

Future 5G New Radio (NR) networks will support a variety of connections with heterogeneous requirements, built upon three generic types of connectivity illustrated on Figure~\ref{fig:scenario3}: extended mobile broadband (eMBB), ultra-reliable low-latency communication (URLLC) and massive machine-type communication (mMTC).
eMBB is a natural unfolding of LTE, where the primary goal is to increase the user data rates and network spectral efficiency.
The enhancements provided by eMBB target mainly human-type traffic, such as high-speed wireless broadband access, ultrahigh-quality video streaming, Virtual Reality and Augmented Reality.

The other two services, URLLC and mMTC, are the cornerstones of machine-type traffic and thus enablers of various types of Internet of Things (IoT) connectivity. 
%foreseen and enabling yet-unknown applications
URLLC, also known as mission-critical IoT, envisions transmission of moderately small data packets (in the order of tens of bytes) with extremely high-reliability, ranging between $99.999\%$ and $99.9999999\%$, i.e. down to $10^{-9}$ packet error probability \cite{urllc_for_tactile_internet}. 
The user plane latency requirement is most commonly defined to be $\SI{1}{ms}$, including uplink (UL) and downlink (DL) roundtrip transmission \cite{3gpp_38913}. %and may potentially include one retransmission in each direction \EDC{not clear?} \cite{3gpp_38913}.
In URLLC the device activity pattern is often not deterministic, but rather intermittent. However, it is generally assumed that it is likely that only a few URLLC devices that are active simultaneously.
URLLC is seen as an enabler of safety systems, wireless industrial robots, autonomous vehicles (cars, trucks, drones), immersive virtual reality with haptic feedback, tactile Internet, and many others which may not even be foreseen at this point.

mMTC aims to provide service to a vast number of devices, out of which only a certain fraction are active at a given time. 
The packet lengths in mMTC are comparable to URLLC, being assumed to be rather short, in the range of tens of bytes. 
% In mMTC, the latency and reliability are not the pressing constraints and are not particularly enforced. 
The main challenge of mMTC is to enable access for sporadically active devices, such that at any given instant an unknown subset out of the massive set of devices wishes to send messages. 
Most mMTC applications do not have strict delay requirements.  The performance is assessed in terms of how many devices can be served within a certain time-frequency resource grid, assuming a certain predefined level of reliability that is much lower compared to URLLC.
% However, the main challenge of mMTC is then to be able to efficiently serve this large number of devices, from a spectral and energy efficiency perspective.
Spectral efficiency (SE) becomes critical, as the overhead of transmitting very small, sporadic payloads from a large set of devices in a typical LTE-type network becomes considerable \cite{towards_massiveURLLC_short_pkts}.
Energy efficiency is critical at the device side, as a large proportion of the mMTC devices are expected to be low-power sensors that are battery-driven.
Therefore, even if access latency is not critical, collisions occurring in the access will affect the energy consumption of the devices.
Furthermore, the deployment of some of these sensors can be in very remote or in low-coverage areas, which brings up the need of having extended coverage capabilities.
Any massive deployment of connected devices falls into the category of mMTC, where devices are mainly sensors used to gather measurements from various environments, such as weather, industry, energy, agriculture, transport, etc. Note that the term IoT may be encountered as an umbrella term for mMTC, but we emphasize that in the context of 5G, IoT is also extended to cover mission-critical IoT, associated with URLLC. 
% One particular technology that is seen as a prominent enabler of the 5G services is massive MIMO.

\subsection{The Three Services in the Context of Massive MIMO}

A massive antenna array is seen as a distinctive technology in the context of the next-generation wireless systems \cite{five_disruptive_tech_5g,Marzetta2010}.
The large number of spatial degrees of freedom (DoFs) created by the massive arrays contribute to achieving two important features: channel hardening and favorable propagation \cite{massiveMIMOnetworks_book,favorable_propag}.
The occurrence of channel hardening essentially means that the massive multi-antenna pre- or post-processing transforms the channel into almost deterministic quantities, making the links quasi-immune to small-scale variations of the instantaneous channel.
The occurrence of favorable propagation means that as the size of the antenna array grows, the propagation channels to different users become more separable.
Therefore, the many-antenna BS becomes more efficient in mitigating inter-user interference through the  use of spatial multiplexing techniques.

The large array gain, coupled with these two features
%which enable spatial multiplexing of multiple users, 
ensure that massive MIMO is the main enabler of high user throughput and high spectral efficiency for eMBB \cite{massiveMIMO_maximal_SE}, \cite{hoydis_mMIMO_how_many_antennas}. In fact, the majority of works have optimized the use of massive MIMO in the context of the high-rate eMBB service. The use of massive MIMO for IoT connectivity has received significantly less attention in the research literature. Massive MIMO holds a potential to support transmission techniques that are suitable for both URLLC and mMTC. However, the massive MIMO techniques used for eMBB are not directly transferable to URLLC and mMTC due to vastly different requirements of these IoT-related services. %This paper fills this important gap by treating the benefits of massive MIMO, as well as the assoficated tradeoffs, for URLLC and mMTC. 
This paper fills this important gap by identifying the opportunities, challenges and tradeoffs of exploiting massive MIMO for URLLC and mMTC.

In URLLC,  the spatial DoFs are essential in attaining the high reliability requirements, as the low-latency requirement severely reduces the time diversity.
% Due to the low-latency feature of URLLC, time diversity disappears. Therefore, the frequency and spatial DoFs need to be efficiently utilized in order to guarantee the strict requirements.
The large number of spatial DoFs contribute to the channel hardening, reducing the odds of having a low energy channel realization that might result in an outage event. Furthermore, the large number of spatial DoFs combined with favorable propagation, enable good spatial separation of devices and thereby efficient spatial multiplexing. However, the efficient use of the large number of spatial DoFs is critically dependent on the channel estimation process, which in massive MIMO can be time-consuming and therefore not directly suited for low latency services. Thus, in supporting URLLC, massive MIMO operation should depart from its eMBB-optimized operation, in which the objective is to estimate the channel as fully as possible and then load the spatially multiplexed channels with highest possible rate. In URLLC, low latency dictates a shorter estimation period, which means that massive MIMO should either operate in a non-coherent mode or shorten the estimation process by relying on a structure present in the channel. Such structure can be manifested through, fro example, the second-order statistics. Furthermore, imperfect channel estimation affects inter-user interference, such that sometimes Time-Division Mutliple Access (TDMA) may be preferable to imperfect spatial multiplexing of users. 
Channel estimation complexity in the case of massive MIMO has lead also to a preference towards time-division duplexing (TDD), since in this case one can rely on the channel reciprocity between UL and DL channels. This is in contrast to frequency-division duplexing (FDD), where the task of estimating the DL channel becomes tedious.

The requirements of mMTC pose an entirely different set of problems, such as activity detection \cite{Polyanskiy2017,Liu2018,Haghighatshoar2018new, Fengler2019}, collision resolution \cite{Bjornson.2017, Han.2017, Marinello2019, JCM_TA_RD_PP_EdC2019} and wide-area low-rate coverage.
Massive MIMO can be used to efficiently support mMTC due to the following two features: \emph{(i)} the favorable propagation property enables multi-user detection (MUD) for a large number of devices; \emph{(ii)} the array gain can boost the signal-to-noise ratio (SNR) for extended coverage requirements. Considering the fact that the central problem in mMTC is the detection of the activity and decoding of the data for an unknown subset of sporadically active devices, the primary benefit from massive MIMO for mMTC is likely to be in the large spatial multiplexing capability. The large number of antennas are instrumental in utilizing the sparsity in activity detection. Furthermore, in the event of sudden, large number of correlated arrivals, the large array is capable to absorb and resolve a large number of devices as well as decode the associated data packets.  

This paper addresses the means through which massive MIMO enables URLLC and mMTC. Besides the support of the individual services, we also discuss the suitability of massive MIMO to support a mix of different services through slicing of the wireless resources~\cite{Popovski2018}. This will reveal interesting tradeoffs that arise due to interaction among the services while trying to exploit the massive number of antennas. As this paper shows, massive MIMO can be a powerful instrument in the quest of fulfilling the requirements of IoT connectivity. 
However, in order to fully reap its benefits, physical layer signal processing techniques need to be integrated with protocol design, with techniques such as channel estimation, retransmission, random access, scheduling, etc.

The paper is organized as follows. The next section discusses the related work. Section~\ref{sec:3gpp} discusses 5G standardization aspects and the 3GPP approach to MTC and mMIMO. 
Sections~\ref{seq:urllc}~and~\ref{seq:mMTC} discuss specific aspects of employing mMIMO to achieve URLLC and mMTC, respectively.
Section~\ref{seq:network_slicing} provides considerations on network slicing in a mMIMO network, and Section~\ref{seq:concl} provides conclusions and offers further research directions.

\section{Related works} \label{sseq:related_works}

\subsection{Research Literature}

\subsubsection{Massive MIMO for eMBB}
%\EDC{cite Marzetta + Erik. Then, state the diverse aspects that have been tackled referencing papers. Do not introduce papers and speak about what they are going. }\\
%\EDC{a higher level view is necessary. Two main metrics: spectral efficiency and energy efficiency. Multi-cell systems. Pilot contamination has become a prominent topic of research. Trade-offs: training/data etc. Analysis with more sophisticated channel models still showing advantages of massive MIMO. }\\
The quest for maximizing SE in eMBB is considered to be the main motivation of using massive MIMO \cite{Marzetta2010,massiveMIMO_maximal_SE}. In order to achieve high SE, careful optimization of the number of spatially multiplexed users, as well as channel training overhead needs to be considered.
% The authors of \cite{massiveMIMO_maximal_SE} explore the trade-off between pilot overhead and data symbols within a coherence interval, and treat the optimal number of users that should be served with a given number of BS antennas such that the SE is maximized.
As the number of antennas increase, channel hardening becomes more effective \cite{Marzetta16}. On the other hand, as the channel exhibits more spatially correlated scattering, the channel hardening effect occurs to a smaller extent \cite{massiveMIMOnetworks_book,Marzetta16}. Therefore, the trade-off between array size and the number of DoFs per user has been investigated \cite{hoydis_mMIMO_how_many_antennas} from the perspective of the achievable rate.
% The question of how many antennas are required is posed in \cite{hoydis_mMIMO_how_many_antennas}, where the authors consider imperfect channel state information, pilot contamination, and a spatial correlated channel model. The main metric of the paper is the achievable rate, and the paper shows an interesting trade-off which appears between the SNR and the spatial DoFs per user. 

Pilot contamination in a multi-cell massive MIMO system is largely considered as the most severe limiting factor in achieving high SE \cite{ach_rates_DL_pilot_contam, superimp_pilot_contamination,superimp_pilots_SE_EE}. However, recent developments \cite{massiveMIMO_unlimited_capacity} show that even in the presence of pilot contamination, the capacity of massive MIMO systems is unbounded as the number of antennas grow, for the case of spatially correlated channels.
\subsubsection{URLLC and massive MIMO}
% \EDC{Is the SoA specific to massive MIMO, if so tell it} 
The general principles of achieving URLLC have been discussed in \cite{URLLC_principles_magazine}, where the importance of optimizing the reliability of the signaling information is highlighted, as well as efficient utilization of the diversity sources.
URLLC is treated from a massive MIMO perspective in \cite{massive_MIMO_tactile_internet}, where the authors emphasize the requirements of tactile Internet and how many antennas are needed to meet the requirements for a given number of simultaneous users, using maximum-ratio transmission (MRT) and zero-forcing (ZF) beamforming, and assuming ideal channel estimation. However, in practice, channel estimation is rarely ideal, being affected by imperfections or pilot interference between users. In this sense, \cite{Wang2016} proposes exploiting the forward error correction (FEC) code diversity in order to assign unique user signatures, which allow separating the pilot-interfering users.

Coherent and non-coherent UL transmissions are compared in \cite{feasibility_large_arrays_urllc} for the case of multi-antenna base-station (BS) and single- or two antennas at the transmitting device. The authors conclude that non-coherent transmission requires higher SNR to achieve the target reliability, and that the single-antenna at the transmitter is not a limiting factor in achieving high reliability.

A stochastic network calculus approach is adopted in \cite{delay_performance_miso_comm}, where the authors evaluate the latency-reliability trade-off using the delay violation probability. It is shown that increasing number of antennas (however, only up to $M=10$) considerably reduces the delay violation probability. A similar approach is taken in \cite{urllc_mmwave_massivemimo} with respect to the evaluated metric, where the authors model the latency-reliability trade-off by imposing a probabilistic constraint on the length of the queue at the BS. 
The end-to-end delay, comprising queueing delay, frame transmission and backhaul latency has been considered in \cite{energy_eff_RA_maMIMO_URLLC}, where the authors consider the joint optimization of transmit power, bandwidth and number of active antennas for a given number of active users in order to maximize the energy efficiency under Quality-of-Service (QoS) constraints of a massive MIMO network.

\subsubsection{mMTC}
% Related Works include: 
In \cite{Bockelmann2018}, authors focus on the massive mMTC service within a multi-service air interface. 
Authors discuss in a broad sense the different physical and medium access techniques to tackle the problem of a massive number of access attempts in mMTC. 
The main conclusion is that, in order for mMTC to take place, there is a need for efficient access protocols capable of withstanding a massive number of devices that contend for network access. 
Physical layer techniques include a) multi-user detection using compressive sensing (CS) techniques, b) collision resolution and harnessing of interference using physical layer network coding, and c) non-orthogonal access with relaxed time-alignment. 
In terms of medium access layer, techniques include a) coded random access and signature based access, b) one/two-stage random access and fast uplink access protocols with a focus on latency reduction.

Grant-free mMTC activity and data detection is tackled in \cite{GF_mMTC_maMIMO_CS}, where the authors consider a massive MIMO scenario and employ compressed sensing to retrieve the device activity and the short messages.
Another approach of solving the activity detection and collision resolution is the grant-based \textit{strongest-user collision resolution} (SUCR) protocol described in \cite{Bjornson.2017}. 
%\EDC{the main reason is to have a transmission that is interference free in the training phase.}
The reason for using such an approach is to be able to solve the collisions prior to the channel training phase and data transmission, with a limited number of orthogonal pilot sequences which is much lower than the total number of devices.
% that the number of orthogonal pilot sequences is limited by the channel coherence time. 
In an mMTC scenario, the number of devices can easily exceed the number of orthogonal pilots; therefore, collision resolution methods have been proposed in the literature, and are summarized in \cite{mag_RA_maMIMO}. 

Throughout the paper we will use boldface small ($\mathbf{x}$) and boldface capital letters ($\mathbf{X}$) to denote vectors and matrices, respectively. 
The superscripts $(\cdot)^*$, $(\cdot)^T$ and $(\cdot)^H$ denote the conjugate, transpose and the conjugate transpose operations, respectively, and $\abs{\cdot}$ and $\vectornorm{\cdot}$ denote the absolute value and the $\ell_2$ vector norm.
The notations $\Pr[\cdot],\mathbb{E}[\cdot]$ and $\text{RSD}[\cdot]$ denote the probability of an event, the expectation and the relative standard deviation of a random variable, respectively.

%%%%%%%%%%%%%%%%%%%%%%%%%%%%%%%%%%%%
\subsection{3GPP standardization \label{sec:3gpp}} %\colnb{Beatriz?}
%%%%%%%%%%%%%%%%%%%%%%%%%%%%%%%%%%%%
5G NR will include the key aspects of massive MIMO, namely advanced antennas with complex digital beamforming, or hybrid analog/digital beamforming, and large antenna arrays. In 2018, 3GPP froze the first 5G NR specification, Release 15 \cite{TS38.300}, focusing on early commercial deployments and a subset of the 5G requirements, mainly related to eMBB. 5G NR R15 supports a maximum of 256 antenna elements, compared to the 64 elements in Release 13. With massive MIMO, beamforming is exploited by combining multiple antenna elements to focus the power in a specific direction. 5G NR specifies new initial access techniques for beamforming that will utilize beams sweeping so that the BS can identify the strongest beam and establish a connection. The MIMO implementations in NR support frequencies below and above 6~GHz, and FDD and TDD.

The second phase of the standardization, to be finalized at the end of 2019, targets fulfillment of the full set of 5G requirements. Particularly, there is a work item dedicated to the MIMO improvements that includes, among others, enhancements to the multi-beam and multi-transmission point operation and to the multi-user MIMO (MU-MIMO). Physical layer enhancements for URLLC and the industrial IoT are also part of the on-going work. 

One remarkable feature for the support of massive MIMO was the introduction, already in LTE-A, of Full-Dimension MIMO (FD-MIMO). FD-MIMO utilizes an active antenna system (AAS) with a 2D plannar array structure, which allows for a large number of antenna elements to be packed. Even more important is the possibility of adaptive electronic beamforming, to form a beam in both horizontal and vertical direction and cover any point in the 3D space.  

\begin{figure}[t]
    \centering
    \includegraphics[width=1\linewidth]{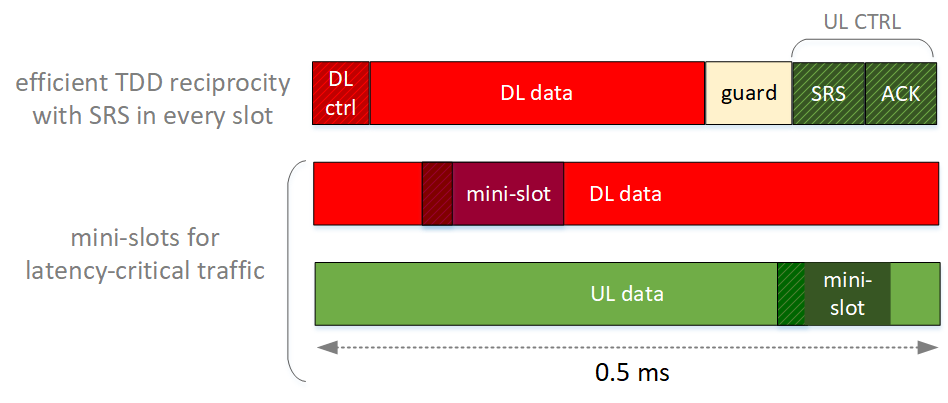}
    \caption{Examples of 5G TDD frames. }
    \label{fig:3gpptdd}
\end{figure}

The 5G NR frame \cite{TR38.802} has been designed with the premise of providing the necessary flexibility to support a heterogeneity of 5G services and requirements. Figure \ref{fig:3gpptdd} illustrates some possible configurations. The general design principle is that static and/or strict timing relations are avoided. For example, asynchronous hybrid automatic repeat request (HARQ) is used instead of predefined retransmission time. A slot is defined as 7 or 14 orthogonal frequency-division multiplexing (OFDM) symbols for subcarriers up to 60 kHz, and 14 OFDM symbols for subcarrier spacing higher than 60 kHz. Data transmission can span multiple slots, to increase the coverage or reduce the overhead due to switching and control information. The TDD DL/UL scheme is much more flexible than in LTE: a slot can contain all DL, all UL, or almost any other DL/UL ratio, and the pattern can be changed in each slot or subframe. A faster TDD switching allows for a more flexible capacity allocation. 
The possibility of having sounding reference symbols (SRS) in every slot allows a more optimized TDD channel reciprocity and therefore more efficient massive MIMO. 
%\PP{Can some of this be illustrated with a figure?}. 
Low-latency for URLLC cases is possible thanks to faster TDD turn-around and the self-contained concept, such that data and ACK can be scheduled in the same slot. For low latency transmissions, a mini-slot has a flexible start position (it can start at any time) and a duration shorter than a regular slot duration. The mini-slot can be as short as one OFDM symbol and it constitutes the smallest scheduling unit. 
%One example of frame configuration with DL and UL traffic is plotted in Figure \ref{fig:5gframe}.  

For the Channel State Information (CSI), 5G NR introduces Type I and Type II CSI \cite{TR38.802}. The former is the normal codebook-based CSI feedback where the device sends back a precoder matrix indicator (PMI) to the gNB, and it supports multi-panel scenarios by having a co-phasing factor across antenna panels. The Type II is an enhanced scheme that enables explicit feedback and/or codebook-based feedback with higher spatial resolution. It can report the wideband and sub-band amplitude information of the selected  beams, such that better precoded MIMO transmission can be employed by the network.

Release 15 has prioritized eMBB services, while the full support for URLLC and mMTC services is expected to be completed in Release 16 and beyond. Specifically, there are on-going activities in the form of work and study items on physical layer enhancements for NR URLLC, channel modelling for indoor industrial scenarios, and NR Industrial Internet of Things (IoT). 

Another important feature in NR is the efficient support of a mix of services. Network slicing is the virtualized technology framework that will accommodate the high requirements heterogeneity of 5G MTC networks \cite{Popovski2018}. Significant progress has been done in 3GPP with regard to the core network and the functional aspects with, e.g., the definition of the network slice identifiers, and procedures and functions for slice selection. However, the exploitation of network slicing in the Radio Access Network (RAN) is not mature in the specification yet. In the research domain, the authors in \cite{Ferrus2019} propose a functional framework for the realization of network slicing management in the RAN, whereas \cite{Popovski2018} discusses the communication-theoretic limits. As it will be discussed in Section \ref{seq:network_slicing}, the implementation of network slicing in a massive MIMO deployment presents additional challenges and opportunities. 

% BS: I have removed this figure because it was too similar to the URLLC frame figure
%\begin{figure}[t]
%    \centering
%    \includegraphics[width=1\linewidth]{figures/5gframe.png}
%    \caption{Example of 5G TDD frame with 30 kHz subcarrier and DL and UL traffic. }
%    \label{fig:5gframe}
%\end{figure}

%%%%%%%%%%%%%%%%%%%%%%%%%%%%%%%%%%%%%%
\section{Massive MIMO solutions for URLLC }
\label{seq:urllc}
%%%%%%%%%%%%%%%%%%%%%%%%%%%%%%%%%%%%%%
% \EDC{insert here the story of the section}
This section presents an overview of the latency and reliability metrics of URLLC, and discusses how the properties of massive MIMO can enable achieving these strict constraints.
Downlink transmission is considered for the most part, as the acquisition of the CSI is more critical at the massive antenna transmitter, especially under a strict latency constraint.
Beamforming methods with imperfect CSI are studied in a single-user TDD system, then extended to the two-user multiplexing case with a varying constraint on the latency.
Feasibility of low-latency FDD operation is also discussed in the DL scenario of URLLC, as well as a few methods for UL transmissions with deteriorated CSI.

\subsection{URLLC Metrics: Physical Layer Reliability and Latency}
In URLLC, the most important metrics are latency and reliability. 
{Several types of latency are encountered in the context of URLLC. 
The most common is the \textit{user-plane radio latency}, defined in \cite{TR38.913,urllc_tcom} as the time duration between a packet arriving at the transmitting layer-2 radio protocol and its delivery at the receiving layer-3 protocol.}
The reliability is defined by 3GPP as the probability of successfully transmitting a number of bytes within a certain user-plane delay. 
More specifically, a general requirement for URLLC is a reliability of ${1-10^{-5}}$ for transmitting a 32-byte packet within 1~ms user-plane latency.
Within the user-plane delay, the user-plane reliability can be increased if retransmissions are performed.
% Therefore, the physical layer latency of a single over-the-air transmission of a packet should be lower than the user-plane latency, thus allowing the user-plane reliability to be increased by means of retransmission.
In this case, the physical layer latency must be smaller than the user-plane latency, whereas the physical layer reliability may be lower than the target user-plane reliability, which would be compensated by retransmissions.

For the remainder of this section, we will only focus on the physical layer latency, which we will simply refer to as latency. 
In a communication theoretic sense, the physical layer latency of a packet is commonly expressed in terms of the number of symbols (or channel uses) it takes to transmit it. 
For a fixed data size, the more symbols available, the lower the rate and the higher the reliability becomes.
Therefore, in a scenario where we have a certain number of bits to transmit, we can aim for a target reliability under varying latency, or for a target latency with unconstrained reliability.
Let us consider the latter option for now, such that $b$ bits need to be transmitted within a certain latency of $N$ symbols. Note that the conversion of the latency into actual time requires specification of the bandwidth user in the system~\cite{urllc_tcom}. Unless explicitly stated otherwise, here we assume a normalized bandwidth and focus on the rate in terms of bits per channel use.
This imposes a transmit rate $R=b/2N$ for a complex channel, where one symbol can be regarded as two channel uses in a real channel.
The physical layer reliability can be expressed as the complementary event to achieving an outage in the transmission, that is ${1-P_\text{outage}}$. 
Considering the fact that the impact of quasi-static fading dominates the effect of the finite blocklength \cite{towards_massiveURLLC_short_pkts}, the outage probability is defined as follows: 
\begin{eqnarray}
P_\text{outage} = \Pr \left[ R > \log_2 (1+ \gamma ) \right]. \label{eq:Pout}
\end{eqnarray}
Provided that the rate is fixed, as well as that transmissions are scheduled and are thereby free of interference, from \eqref{eq:Pout} it follows that there exists a threshold SNR $\gamma_\text{th}=2^R-1$ such that a packet encoded with rate $R$ can be reliably decoded. Therefore, the outage probability can be expressed as:
% \EDC{you do not need this definition if you have (1). It is the same. But you can define $gamma_th$}
\begin{eqnarray}
P_\text{outage} = \Pr\left[\gamma<\gamma_\text{th}\right].
\end{eqnarray}

\subsection{Favorable Massive MIMO Properties and CSI Acquisition}
Due to the low-latency constraint, it is very challenging to use time diversity in order to support URLLC. At the physical layer URLLC can benefit from frequency diversity provided by the wideband OFDMA, as well as from space diversity offered by the massive antenna arrays.
Space diversity manifests through the two phenomena mentioned previously: channel hardening and favorable propagation \cite{massiveMIMOnetworks_book,Marzetta16}. Channel hardening is particularly important for URLLC, as it diminishes the impact of small-scale fading in a similar fashion as time diversity would do over multiple coherence intervals.
Favorable propagation is important when performing spatial multiplexing of several URLLC devices, as it ensures that the streams to the devices can be well separated at the massive array BS, provided that the BS has obtained CSI from the devices.
% \EDC{benefitting from favorable conditions: we have to have the CSI}

%\subsection{CSI acquisition in massive MIMO}

%\EDC{You have to say somewhere that we study downlink and why. I suggest at the beginning of section III, you explain the content and motivation of the section}

In order to make use of downlink precoding schemes, the massive array transmitter must be aware of the CSI.
The task of acquiring DL CSI at the transmitter is challenging in FDD systems, since the training length is dependent on the number of BS antennas.
In contrast, in TDD systems channel reciprocity is assumed between UL and DL, and therefore, UL training can be utilized to estimate the channel coefficients at all antennas simultaneously, i.e. the training length is independent of the number of antennas.
%\EDC{say more precisely why. The text should be self contained.}
This mode of operation alleviates the CSI acquisition procedure, such that massive MIMO is commonly based on TDD operation. 
%In TDD operation, the devices send UL pilots, from which the massive array BS estimates the channel, then utilizes it for DL precoding.
\begin{figure}[t]
    \centering
    \includegraphics[width=1\linewidth]{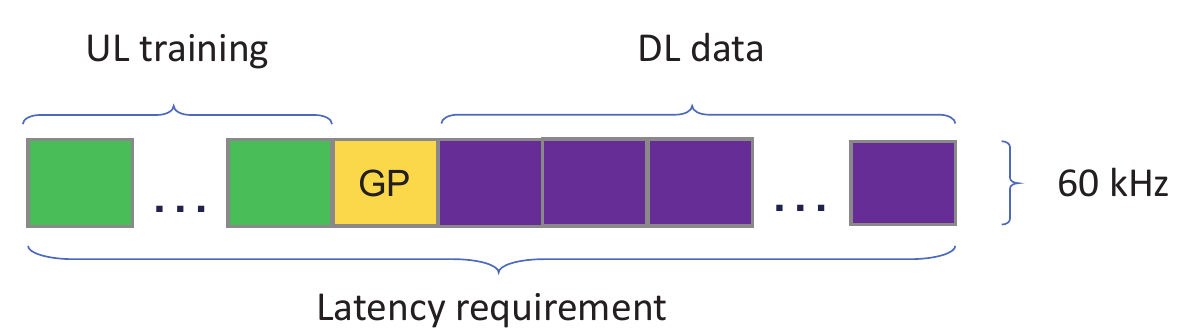}
    \caption{An URLLC frame containing UL training symbols, guard period (GP) and DL data, with 60 kHz subcarrier.}
    \label{fig:frame_urllc}
\end{figure}
\subsection{Imperfect full CSI}
Assuming the case of one single-antenna device, the BS can estimate the channel coefficient for each antenna based on the UL pilots sent by the device by using the least squares (LS) method. The $M\times 1$ vector estimate of the channel coefficients can be expressed as
% The received pilot signal at the BS can be expressed as the $M$ by $t$ matrix 
% \begin{eqnarray}
% \mathbf{Y}_\text{UL,p} = \mathbf{h} \mathbf{p}+ \mathbf{N}, 
% \end{eqnarray}
% where $M$ is the number of BS antennas, $t$ is the number of pilot symbols per device, $\mathbf{h}$ is the $M\times1$ channel vector, $\mathbf{p}$ is the $1\times t$ pilot vector fulfilling $\mathbf{p}^H \mathbf{p}=t\mathbf{I}$, and $\mathbf{N}$ is a  $M\times t$ matrix with the complex circularly symmetric zero-mean Gaussian noise, where the columns are i.i.d. realizations of $\mathcal{CN}(0,\sigma_n^2\mathbf{I})$.
%\EDC{which quantity does this refer to: a vector? which vector?}.
% We assume the BS performs least-squares (LS) estimation of the channel coefficients, which can be expressed as
\begin{eqnarray}
\widehat{\mathbf{h}} = \dfrac{1}{t} \mathbf{Y}_t \mathbf{p}^H
\end{eqnarray}
where $t$ is the training length, $\mathbf{p}$ is the $1\times t$ pilot vector fulfilling $\mathbf{p}^H \mathbf{p}=t\mathbf{I}$, and $\mathbf{Y}_t$ is the $M\times t$ received training signal.
%\EDC{I guess there are some assumptions on p to be able to write that}
The noisy LS estimates for the channel coefficients at each BS antenna $\widehat{\mathbf{h}}$ are then used to perform DL MRT.
The transmitter uses the precoding filter $\mathbf{w}_\text{MRT}=\widehat{\mathbf{h}}^H/\vectornormsmall{\widehat{\mathbf{h}}}$, and the DL MRT SNR at the device is then:
\begin{eqnarray}
\gamma^\text{DL}_\text{MRT}=\dfrac{\rho}{\sigma_n^2} \dfrac{\abs{\widehat{\mathbf{h}}^H \mathbf{h}}^2}{\vectornormbig{\widehat{\mathbf{h}}}^2}, \label{eq:SNR_MRT}
\end{eqnarray}
where $\rho$ and $\sigma_n^2$ denote the transmit power and the noise variance, their ratio being the pre-processing SNR.
%\EDC{replace $\sigma_h$ by transmit power.}

\subsection{Exploiting channel sparsity}
%\EDC{say why we examine this case}
In reality, fading channels are not seen as i.i.d. coefficients from the massive array BS. 
Instead, the coefficients experience a certain correlation, based on the spatial structure of the propagation environment.
Let us consider the case of a cluster-based channel model, in which each cluster is characterized by a set of propagation paths based on their angle of departure (AoD) and their angle of arrival (AoA) \cite{massiveMIMOnetworks_book,hoydis_mMIMO_how_many_antennas}.
In addition, each propagation path incurs an attenuation given by an independent fading coefficient following a zero-mean circularly symmetric complex Gaussian distribution, with exponentially distributed power, decreasing with a power decay factor, defining the maximum decrease between the strongest and weakest path.

In a simplified form, for a single-antenna user, the channel can be expressed as the following column vector
\begin{eqnarray}
\mathbf{h} =  \sum_{i=1}^{N_P} \alpha_i \mathbf{s}_{i}. \label{eq:h_np}
\end{eqnarray}
Here, $N_P$ is the total number of paths, $\alpha_i$ is the fading coefficient of path $i$, and $\mathbf{s}_{i}$ is the normalized steering vector for each path. %\EDC{well, this is a simplified cluster based model.}
%\EDC{distribution of alpha's}\asb{mentioned in the paragraph before}

This spatial structure of the channel can be captured using second order statistics, more specifically, by estimating the channel correlation matrix of each device at the BS \cite{massiveMIMOnetworks_book}, which is defined as $\mathbf{R}=\mathbb{E}[\mathbf{h}\mathbf{h}^H]$, where the expectation is taken over the channel realizations.
%\EDC{expectation wrt which quantity}
In addition to the TDD channel reciprocity, let us assume that the 
the correlation matrix of the device is perfectly known at the BS. This sets the context to discuss how the precoding can be enhanced using the second-order statistics of the channel.

The correlation matrix takes the following form when expressed based on the singular value decomposition:
\begin{eqnarray}
\mathbf{R}= {\mathbf{V}} \Lambda {\mathbf{V}}^{H}.   \label{eq:cov2}
\end{eqnarray}
Here, $\mathbf{V}$ is an $M\times N_p$ matrix containing the singular vectors of the channel and $\Lambda$ is the diagonal matrix containing the $N_P$ eigenvalues of the channel.
Both the instantaneous estimated channel coefficients $\widehat{\mathbf{h}}$ and the channel second-order statistics in the form of the singular vectors (SV) $\mathbf{V}$ can be utilized in forming the DL precoding vector $\mathbf{w}_\text{SV}$.
%\EDC{ Not clear, to be improved.}

The procedure \cite{asilomar_urllc_maMIMO} consists in projecting the instantaneous channel estimate vector $\widehat{\mathbf{h}}$ on the subspace spanned by the singular vectors $\mathbf{V}$.
The result is $\widehat{\mathbf{h}}^T \mathbf{V}$, which represents the instantaneous fading coefficients for each singular vector, thereby being a refined instantaneous estimate, as the noise lying outside of the subspace of the singular vectors is eliminated.
This result is then used to form the projected estimated channel $\widehat{\mathbf{h}}^T \mathbf{VV}^H$, its matched filter being the beamforming vector

%\EDC{add that we could emply some compressed sensing methods but too complex and takes too much time in view of the latency}
% thereby the removing the channel estimation noise that lies outside of this subspace.
% The result of this first projection is the estimated instantaneous channel for each singular vector, which is then projected onto the subspace of $\mathbf{V}$ and match-filtered in order to obtain 
\begin{eqnarray}
\mathbf{w}_\text{SV}= \dfrac{\mathbf{VV}^H\widehat{\mathbf{h}}^*}{\vectornormbig{\mathbf{VV}^H\widehat{\mathbf{h}}^*}}.
\end{eqnarray}
The DL SNR has then been showed \cite{asilomar_urllc_maMIMO} to be
\begin{eqnarray}
\gamma^\text{DL}_\text{SV} = \dfrac{\rho}{\sigma_n^2} \dfrac{\abs{\mathbf{h}^T\mathbf{VV}^H\widehat{\mathbf{h}}^*}^2} {\vectornorm{\mathbf{VV}^H\widehat{\mathbf{h}}^*}^2}. \label{eq:SNR_SV}
\end{eqnarray}
Compressed sensing methods could potentially be employed as well, however, at the expense of higher complexity and latency. 

\subsection{Impact of training duration}
\begin{figure}[t]
    \centering
    \includegraphics[width=1\linewidth]{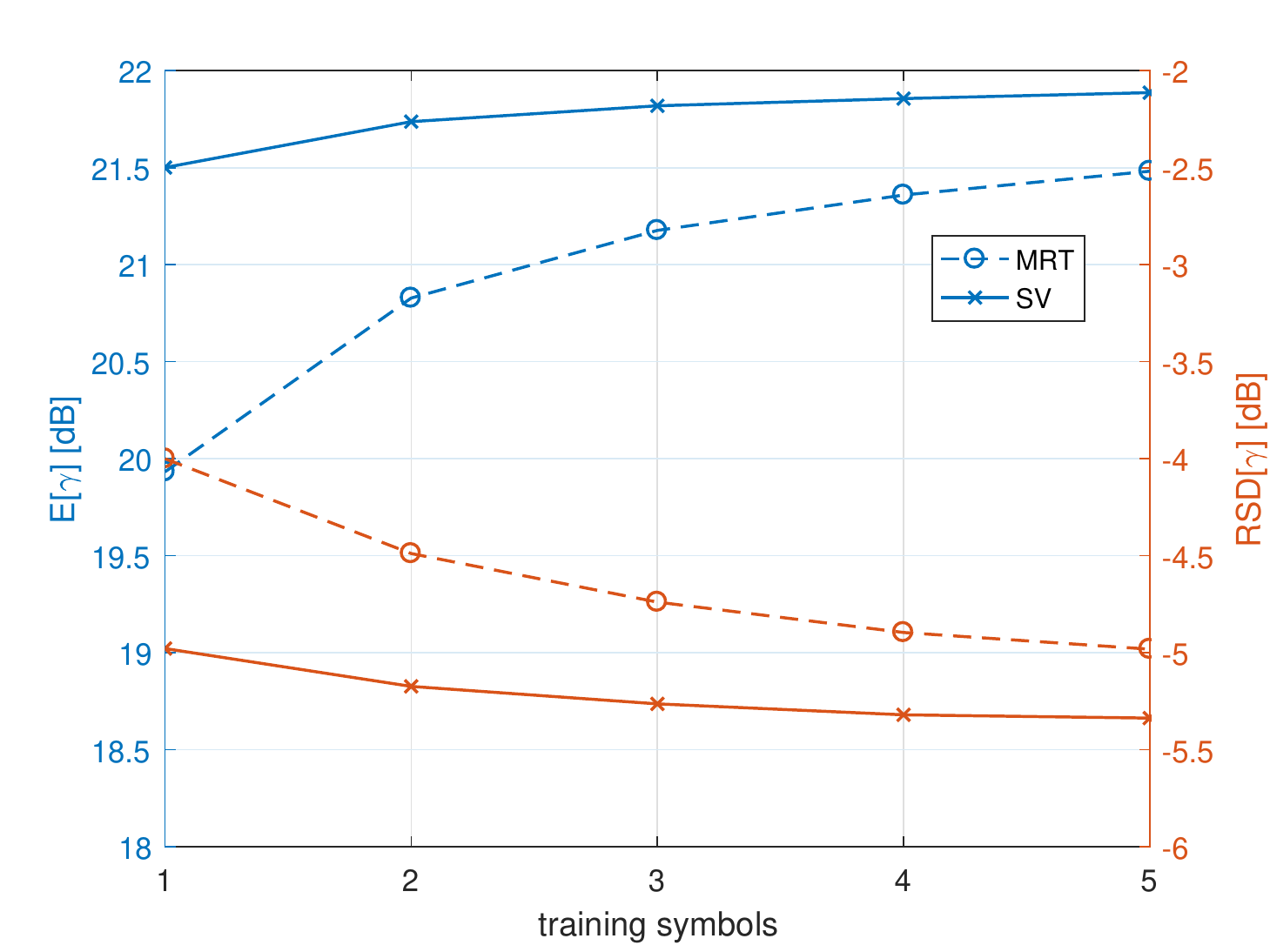}
    \caption{Dependency of the mean and relative standard deviation of the DL SNR $\gamma$ on the training length. The pre-processing SNR is 4.5~dB and the channel consists of 20 paths with exponential decay of up to 10~dB.}
    \label{fig:snr-training}
\end{figure}

\begin{figure}[t]
    \centering
    \includegraphics[width=1\linewidth]{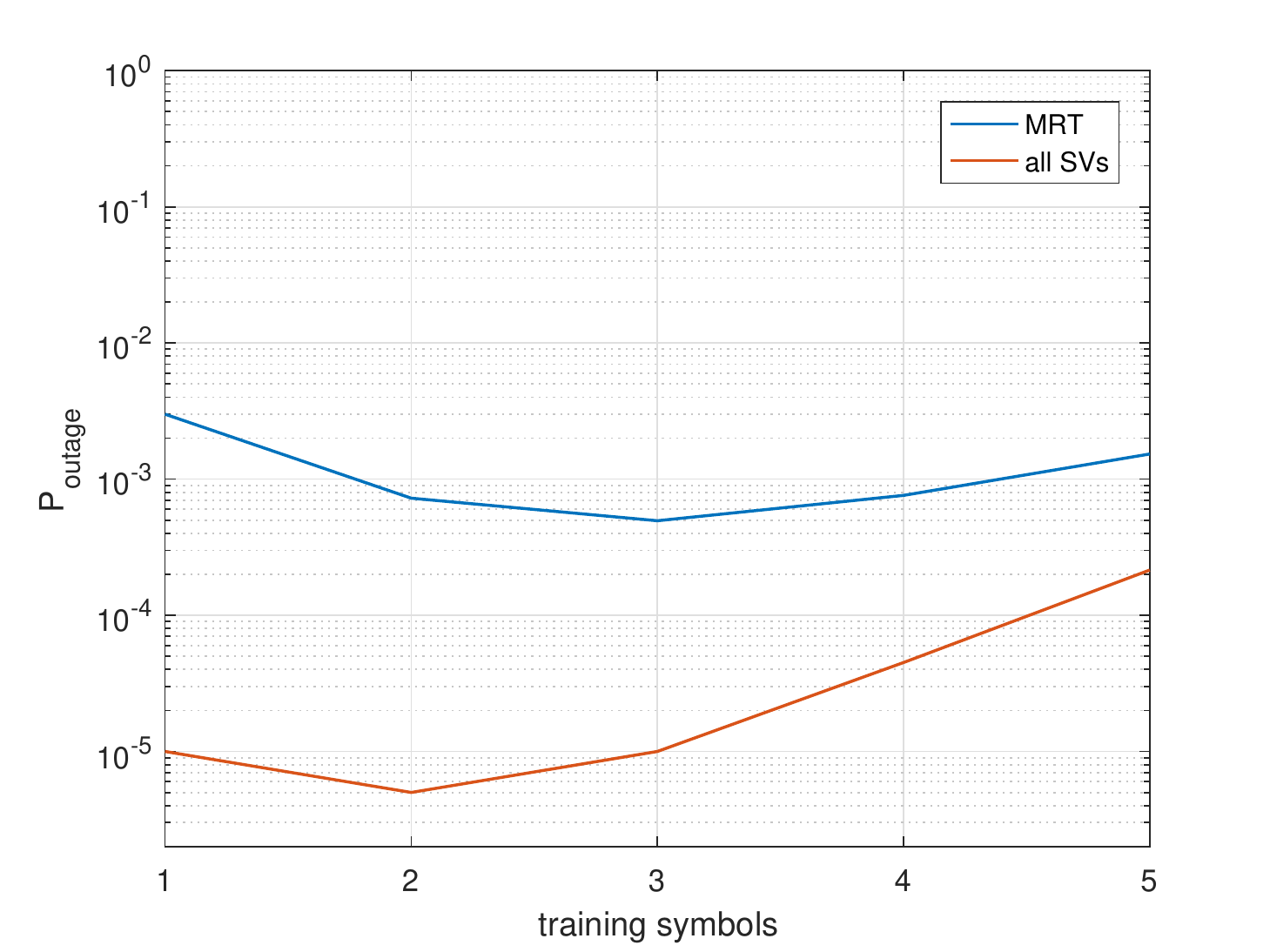}
    \caption{Dependency of the outage probability on the training length. The pre-processing SNR is 4.5~dB and the channel consists of 20~paths with exponential decay of up to 10~dB. SCS of 60~kHz, 2~slots of 14~symbols, comprising 28~symbols in total, with a duration of 0.5~ms.}
    \label{fig:outage-training}
\end{figure}

The training length is an important parameter, especially when dealing with low-latency transmissions.
Figure~\ref{fig:snr-training} shows the dependence of the mean and relative standard deviation (RSD) of the DL SNR $\gamma$ on the training length, for the MRT scheme and for the SV-based precoding.
It can be seen that increasing the number of training symbols for the SV-based scheme provides minimal increase in average SNR and minimal decrease in RSD. 
However, it can be noticed that MRT is more sensitive to the training length, such that increasing the training length can be beneficial.
Figure~\ref{fig:outage-training} shows how the outage varies in a latency constrained scenario where the total number of channel uses are limited, as the training symbols are increased.
It can be seen that for the case of 28 symbols (2 slots of 14 symbols with one subcarrier of 60 kHz, corresponding to 0.5 ms duration), the optimal number of training symbols is slightly different for the two schemes.
Note that as more symbols are used for training, fewer symbols remain for data, therefore, a higher rate and SNR threshold being required to successfully decode the packet.
Furthermore, it can be seen that the scheme relying on projecting the channel estimate on the singular vector subspace achieves between 1 and 2 magnitudes lower outage probability.
It is, therefore, worthwhile to consider such a precoder which can exploit the channel correlation in order to refine the instantaneous CSI.

\subsection{Multi-user setting: SDM vs TDM}
In a multi-user setting, another trade-off arises in terms of how to allocate the DoFs between the multiplexed devices.
If spatial multiplexing, or space division multiplexing (SDM), is employed, the spatial DoFs are shared, whereas if time division multiplexing (TDM) is employed, the devices use the full spatial DoFs, but share the time resources.

SDM can be particularly suitable when the devices share the same latency requirements, as they can be served simultaneously with a fixed physical layer latency. Therefore, in SDM, the average device latency is equal to the system level latency. Here, by system level latency we consider the total latency required to serve the devices.

TDM, on the other hand, may be suitable when the latency requirements may differ, or when traffic is prioritized. In TDM, the devices experience different latency, depending on the transmission priority. Therefore, the average device latency is smaller than the system level latency.

\begin{figure}[t]
    \centering
    \includegraphics[width=1\linewidth]{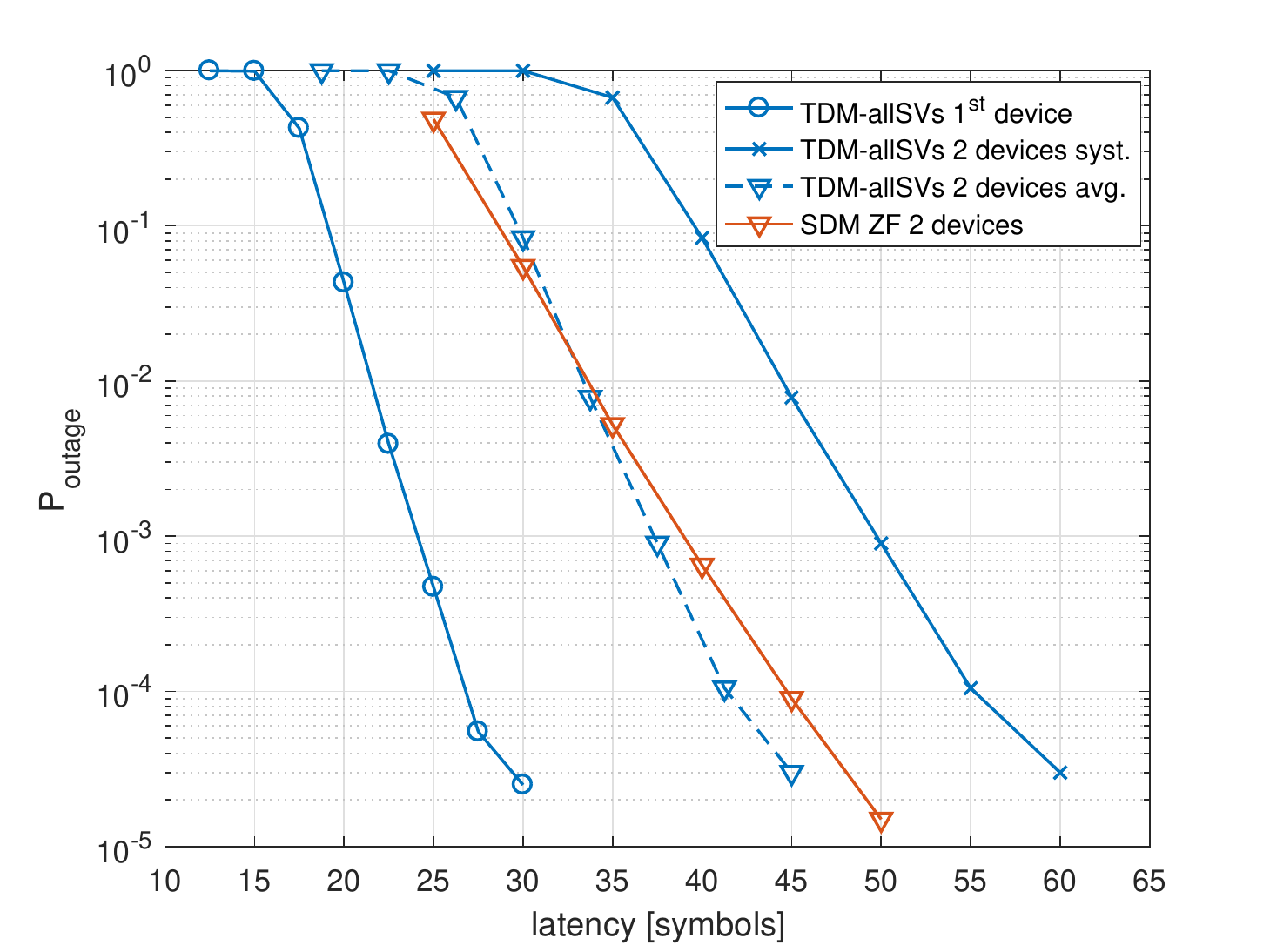}
    \caption{Comparison between latency-outage performance of TDM and SDM for the DL transmission to two devices. TDM relies on using the refined instantaneous estimate over the SVs of the channel covariance matrix, whereas SDM relies on ZF based on instantaneous channel estimates.
    The dashed line indicates the average latency-outage characteristic of the TDM scheme. The channel is assumed to be sparse with $N_P=10$ paths, with 10~dB decay and 8~dB pre-processing SNR.}
    \label{fig:2userZF}
\end{figure}

The effects of the two multiplexing strategies can be observed in Figure~\ref{fig:2userZF}. TDM requires higher system latency to serve both devices with the same reliability as SDM with ZF. This is due to the increased data rate caused by the sharing of time resources, which, in turn, causes the threshold SNR $\gamma_\text{th}$ to soar.
However, the first device experiences lower latency and, for this reason, the average device latency is similar for the two schemes, despite the overall system latency being higher in the case of TDM than SDM.
This further states that SDM is more suitable when multiplexing devices with the same latency requirement, and that TDM is suitable for multiplexing devices with distinct latency characteristics.

The efficiency of spatially multiplexing the devices depends on several factors, such as which multiplexing technique is used (here MRT and ZF) and the channel estimation accuracy. 
The signal-to-interference-and-noise ratio (SINR) when multiplexing two devices can be expressed, regardless of the multiplexing technique, as:
\begin{eqnarray}
\gamma_1 = \dfrac{\abs{\mathbf{h}_1\mathbf{w}_1}^2}{\sigma_n^2 \abs{\mathbf{h}_1\mathbf{w}_1} + \abs{(\mathbf{h}_1\mathbf{w}_1)^H \mathbf{h}_1\mathbf{w}_2}^2}
\end{eqnarray}
where $\mathbf{w}_i$ is the beamforming vector for each device, depending on the beamforming method.
If MRT is used, the numerator is maximized, without taking into account the interference term in the denominator.
If ZF is used, the interference term in the denominator is minimized. The extent to which this can be performed is dictated by the accuracy of the channel estimate which is used in $\mathbf{w}_2$. For low-SNR scenarios, several pilot symbols per device or estimation refinement using second order statistics are beneficial in order to improve the multiplexing efficiency.

%\PP{We also need to mention what is the spatial multiplexing efficiency, i.e. how well are different users separated when we rely on this second order statistics. It is partially discussed two subsections later, but would be good to put some expressions for how the users interfere.}

% These effects can be observed in Figure~\ref{fig:2userZF}, where time-division multiple access of two users is compared to space-division multiplexing based on zero-forcing (ZF).
% Multiplexing users with ZF provides better performance, owing to the favorable propagation property of massive MIMO, which enables the BS to spatially separate devices.
% It can be seen from Figure~\ref{fig:2userZF} that ZF achieves 2 orders of magnitude higher reliability than TDMA for the same physical layer latency requirement. \PP{This part needs to be explained. What you are basically saying here is that the use of all symbols by both users, but with halved degrees of freedom per user, is better than using all spatial DoFs for each user, but only for half of the symbols. Is this an artefact of final blocklength or of improved training?}

\subsection{Feasibility of FDD in URLLC} 
Legacy systems benefit from operating in FDD for two main reasons: lower latency and no need for guard-time when switching between UL and DL.
However, due to the channel estimation procedures, massive MIMO is generally assumed to operate in TDD.
FDD operation is possible in massive MIMO only when the channel is assumed to be sparse, such that the BS is able to estimate the instantaneous coefficients of the dominant paths in the DL by transmitting orthogonal pilots, and then receiving the coefficients as a feedback from the device.
This operation incurs considerable overhead, which increases with the number of dominant propagation paths, as the number of DL orthogonal pilots and the number of symbols used for UL feedback scale with the number of dominant paths.

Let us assume that the correlation matrix is known at the BS, therefore the BS knows the dominant singular vectors and their relative power, represented by the diagonal eigenvalue matrix.
With this knowledge, the BS constructs an orthogonal set of pilots equal to the number of singular vectors to be estimated.
Note that the number of singular vectors to be estimated does not necessarily have to be the full size of the dominant eigenspace ($N_s<N_P$).
The DL SNR can be expressed similarly to \eqref{eq:SNR_SV} as:
\begin{eqnarray}
\gamma_\text{FDD} = \dfrac{\rho}{\sigma_n^2} \dfrac{\abs{\mathbf{h}^T\mathbf{V}_{N_s} \widehat{\bm{\beta}}^H}^2} {\vectornorm{\mathbf{V}_{N_s}\widehat{\bm{\beta}}^H}^2},
\label{eq:SNR_FDD}
\end{eqnarray}
where $\widehat{\bm{\beta}} = \widehat{\mathbf{h}_\text{DL}^T\mathbf{V}_{N_s}} $ are the estimated coefficients for each singular vector which are in the simplest case fed back in an analog manner \cite{fdd_massivemimo_analog_feedback}.

\begin{figure}[t]
    \centering
    \includegraphics[width=1\linewidth]{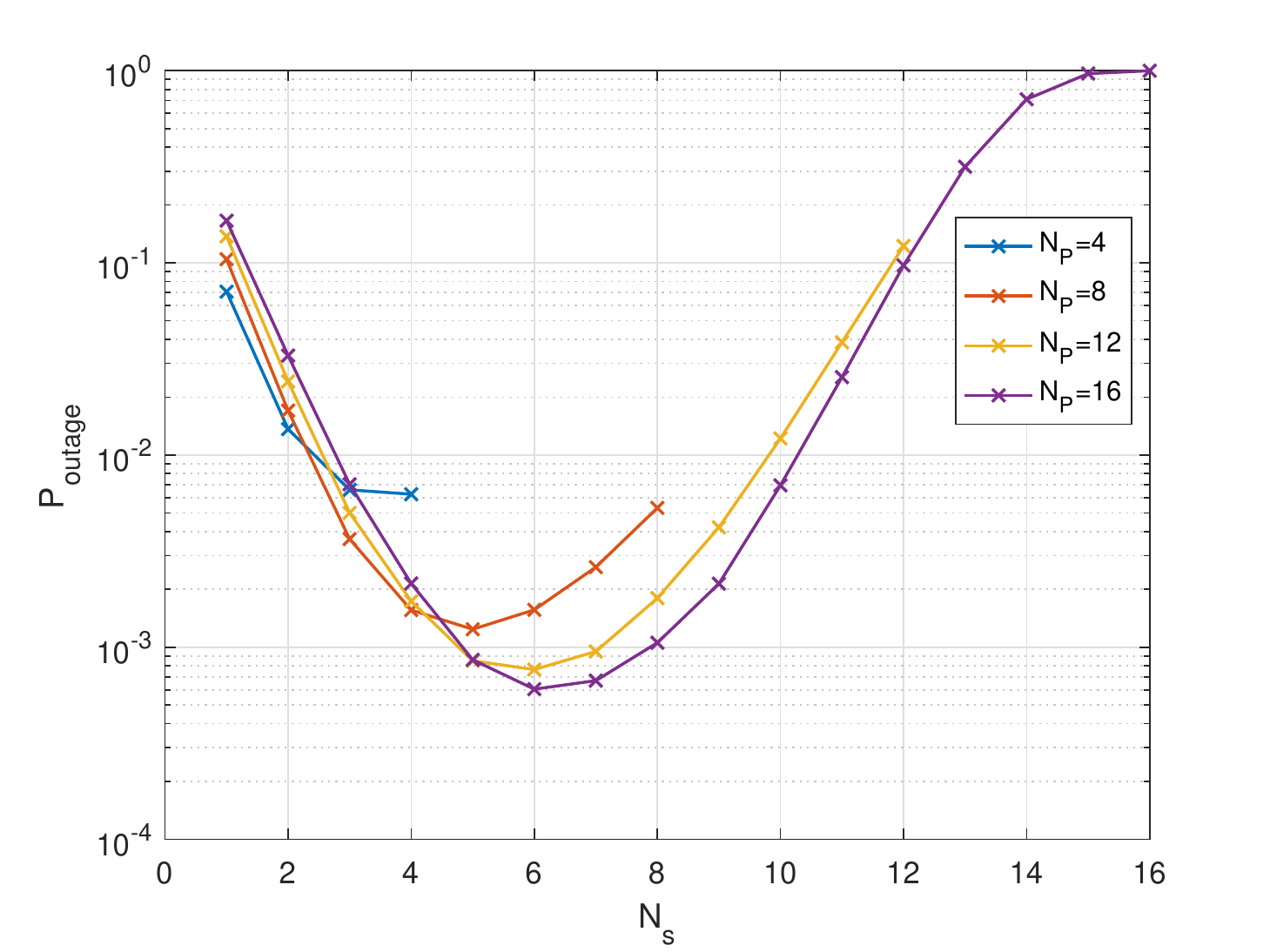}
    \caption{Dependency of $P_\text{outage}$ on the channel diversity ($N_P$) and on the number of pilots invested in estimating $N_s$ instantaneous fading coefficients of the paths. Pre-processing SNR is 10~dB, and the power decay is exponential up to 20~dB.}
    \label{fig:FDD_Ns_Np}
\end{figure}

The trade-off between the outage probability and the number of singular vectors to estimate is shown in Figure~\ref{fig:FDD_Ns_Np}. 
It can be seen that for $N_P=4$, all the singular vector instantaneous coefficients should be estimated in order to minimize the outage. 
However, as the channel diversity increases, the optimal $N_s$ becomes as low as 6 for a channel with a total of $N_P=16$ paths. This is due to the increasing overhead required to estimate additional SVs.

\begin{figure}[t]
    \centering
    \includegraphics[width=1\linewidth]{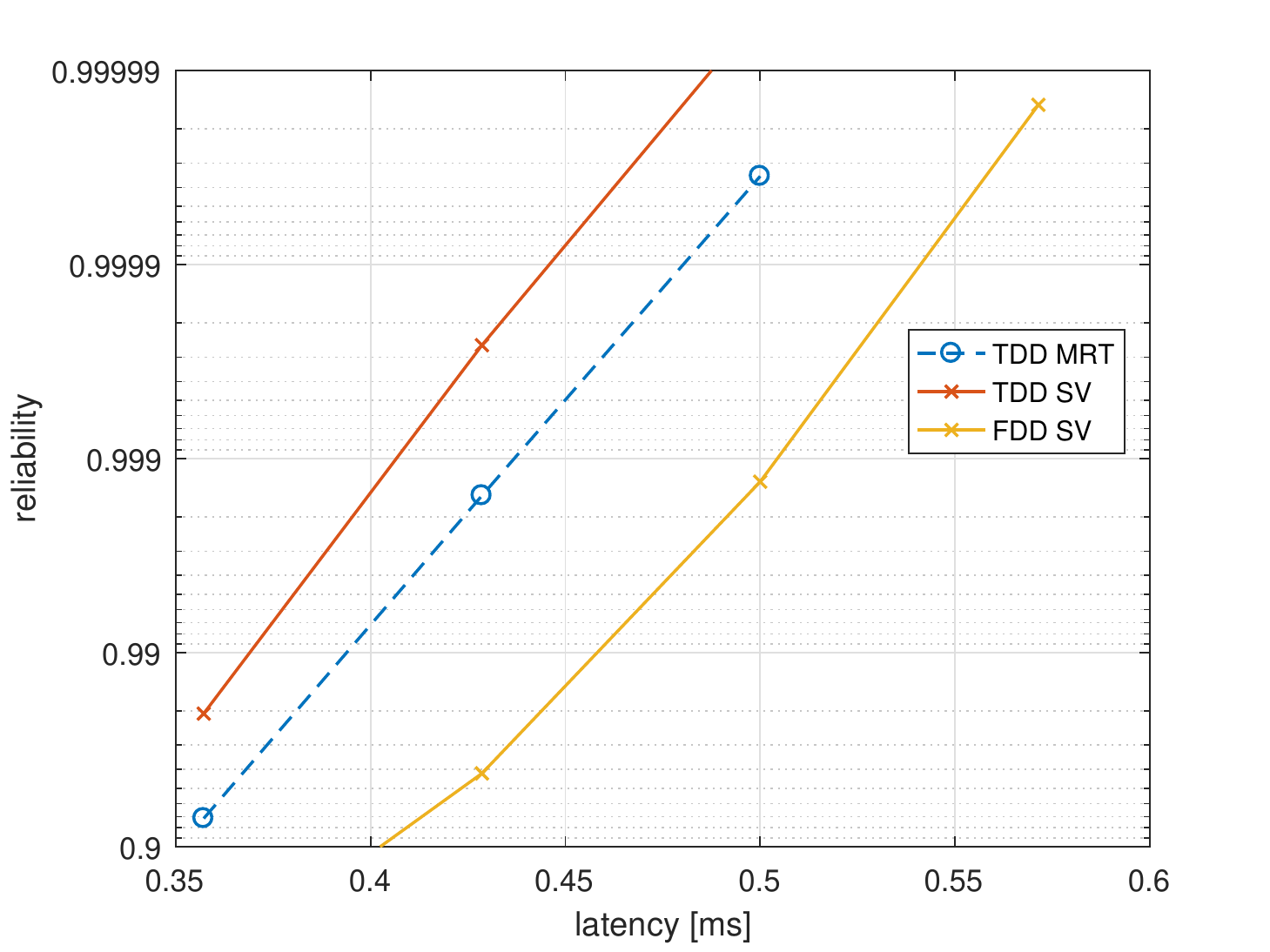}
    \caption{Latency-reliability dependence in the DL using different precoding techniques. The pre-processing SNR is 9~dB and the SCS is assumed 60~kHz. The traffic is assumed symmetric in UL and DL, and the channel is assumed to consist of $N_P=12$ paths with up to 20~dB decay. The number of training symbols is optimized for each latency in order to maximize the reliability.
    % The TDD schemes use 1 UL symbol for training, whereas FDD uses the optimal value $N_s$ which minimizes the outage.
    }
    \label{fig:rel-latencyFDDcomp}
\end{figure}

Figure~\ref{fig:rel-latencyFDDcomp} shows the latency-reliability trade-off for TDD and FDD. 
It can be seen that in TDD, utilizing the SV-based scheme provides considerable improvements over MRT.
For both schemes, the number of training symbols is optimized such that the reliability is maximized for each latency. 
The training varies between 1-3 and 1-2 symbols for TDD-MRT and TDD-SV, respectively, out of a total of between 20 and 40 channel uses defining the latency.
For FDD, the performance is degraded compared to TDD schemes, due to the increased overhead required to estimate the SVs.
The optimal number of DL training symbols that maximizes the reliability for the same 20 to 40 channel uses is varying between 3-9 symbols, being therefore up to 3 times higher than in TDD.
However, this is not to say FDD is infeasible, but only to highlight that in a massive MIMO scenario it becomes less efficient than TDD.
% due to the extra resources spent on training, the performance is degraded compared to the TDD schemes. \asb{more comments on the optimization of training}

% Acquiring CSI at the massive array transmitter (CSIT) is generally more challenging than acquiring CSI at the receiver (CSIR).
% In most scenarios, massive MIMO systems are assumed to be operating in TDD mode, as this assumption eases the CSI acquisition procedure by employing channel reciprocity between uplink and downlink.
% Massive MIMO in FDD mode is restricted by the high complexity CSIT acquisition. Due to the large size of the transmitter array, the CSIT estimation is based on firstly estimating the correlation matrix (long-term statistics) of each user, and secondly on estimating the instantaneous coefficients of each propagation path of each user.
% This procedure functions well for sparse channels, with limited mobility of users. 
% Having rich scattering requires high estimation overhead in order to achieve the full channel diversity, whereas increased mobility of users would require more frequent updating of the channel correlation matrix, which can again become costly.
% \asb{FDD could be attractive for low latency due to feedback}
\subsection{Non-coherent massive MIMO}
For scenarios where the channel may vary rapidly throughout a packet transmission due to high mobility, schemes that do not rely on channel acquisition have been investigated from a reliability perspective \cite{noncoherent_design_performance,dual_stage_non_coh,Baeza2018}.
Non-coherent energy detection (ED) in the UL at a large antenna array has been proposed \cite{noncoherent_design_performance}, as it does not rely on the instantaneous channel coefficients.
The procedure exploits the channel and noise hardening phenomena, which make the channel more deterministic, such that the pulse amplitude modulated signals can be decoded reliably based on the received energy, only with the knowledge of long-term statistics of the channel.
An extension of this idea has been considered in \cite{dual_stage_non_coh}, where a constellation allowing both coherent and non-coherent reception of symbols has been proposed.
The reliability is shown to be increased for a 2~bit/symbol rate compared to the case of QPSK, when channel estimation proves to be inaccurate.
Depending on whether training symbols have been invested in obtaining an instantaneous channel estimate, one of these schemes can be employed in the UL in order to increase reliability in severe fading conditions when CSI is unavailable or degraded.

\section{Massive MIMO for mMTC}
\label{seq:mMTC}
%%%%%%%%%%%%%%%%%%%%%

The challenges brought by mMTC stem from the massive number of deployed MTC devices. 
The MTC traffic is usually sporadic and unpredictable, as the MTC devices transmit in an uncoordinated way whenever they have data to transmit. 
% and a common assumption of mMTC is that only a fraction of the devices would be active at a given time. 
The main question is how to detect the activity and successfully decode the data from a maximal number of transmitting mMTC devices in the uplink within a limited bandwidth.
%It is worth mentioning that some methods presented in the following are also applicable to crowd scenarios, where the devices are user equipments. \\

%\PP{I would remove this part, as finite blocklength is not relevant for all algorithms discussed later on. ||| In this setup, conventional random access solutions assume activation detection of the transmitting devices and the fact that the probability of successful joint decoding goes asymptotically to one with increasing blocklength. However, in the context of mMTC the devices have small data payloads. Even though the subset of devices that are active simultaneously can be rather small, the large total number of devices results in finite blocklength (FBL) effects.}

The general model used for random access in mMTC is that a very large number of devices operate in an uncoordinated fashion. At any time, only a small subset of the total are active, i.e., with some payload to be sent. Massive MIMO plays an important role in mMTC random access. The fundamental benefit offered by massive MIMO is the large spatial multiplexing gain, allowing  accommodation of a large number of devices transmitting simultaneously, resolution of collisions and efficient data decoding, such as compressed sensing. 
Enhanced array gain and channel hardening are important features as well.  In particular, channel hardening enables a simple collision resolution procedure in section~\ref{sseq:grant-based} and grant-free coded-random access transmission in section~\ref{sseq:grant-free}.

\subsection{Random Access Protocols with Massive MIMO}

This section treats several random access protocols models for massive MTC. The protocols differ according to the resource that is randomly accessed, how training is performed, as well as how data is transmitted and decoded. 

\begin{table}[t]
\begin{tabu} to \columnwidth {|X|X|X|X|X|}
\hline
& Pilot sequences & Grant mode & Device ID\\ \hline\hline
SUCRe & Orthogonal & Grant-based & yes \\ \hline
pilot RA  & Orthogonal & Grant-free & yes \\ \hline
 coded RA & Orthogonal & Grant-free & yes \\ \hline
{CS-RA} & Pseudo-random & Grant-free & yes  \\ \hline
Unsourced RA & none & Grant-free & no  \\ \hline
\end{tabu}
% \centering
\caption{Classification of random access protocols according to three features: type of pilot sequence, grant mode and device identification.}
\label{table:RAprotocols}
\end{table}

In order to give an overview of the methods described in this section, we have classified them according to three main characteristics (see table~\ref{table:RAprotocols}):

\begin{itemize}
   
\item \emph{Training:} Most of the methods rely on training and coherent detection. One assumption for the model is that the pilot sequences are orthogonal, but the  devices access the pilot sequences randomly. As the number of orthogonal pilots is limited, there are collisions, meaning that multiple devices may select the same pilot sequence. An alternative solution for training is one based on pseudo-random pilot sequences where a unique sequence is allocated to each user at the expense of decreased channel estimation quality. Finally, pilot-based estimation can be avoided and the system can operate in a non-coherent manner.

 \item \emph{Grant mode:} A commonly used mode is the grant-mode, present e.g. in the legacy 4G systems. The collisions in the pilot domain are first resolved so that each pilot sequence is allocated to a single device and subsequent data transmission occurs collision-free. The other schemes are grant-free and rely on mechanisms for joint training and data transmission without resorting to a preliminary collision resolution phase. 
 
\item \emph{Device identification:} All the methods have a mechanism for device identification at the exception of unsourced random access \cite{Polyanskiy2017}, where the goal is to decode the transmitted messages, without the possibility of  identifying the transmitting devices.
\end{itemize}

\subsection{Grant-based random access protocols}
%=========================
\label{sseq:grant-based}

\subsubsection{SUCRe}
The Strongest-User Collision Resolution (SUCRe)~\cite{Bjornson.2017} protocol is a grant-based random access protocol, described as follows:
\begin{itemize}
\item
\textit{Phase 1.} In SUCRe, the devices accessing the UL randomly transmit a pilot sequence from an orthogonal set of %$\tau_p$
$\mathcal{P}_{p}$ pilots. Due to the large number of devices and the pilot scarcity, several devices may choose the same pilot, resulting in a pilot collision.
\item
\textit{Phase 2.} The BS performs channel estimation based on the orthogonal pilots detected, thereby obtaining contaminated channel estimates for colliding users. Using these estimates, the BS performs precoded transmission. 
\item
\textit{Phase 3.} The devices are able to reliably measure their received signal array gain, owing to the channel hardening effect \cite{Marzetta16}. Thereby, if the array gain is equal to the number of antennas $M$, there has been no collision in the pilot domain and the device is granted access to the pilot sequence. If the individual array gain at the device is a fraction of $M$, a number of devices have collided in the pilot domain. The collision resolution is performed using a distributed decision rule employed at the devices, such that only the strongest user is granted access to the pilot sequence. 
\item
\textit{Phase 4.} The devices that are granted access to a pilot sequence retransmit the pilot sequence  followed by data transmission.
\end{itemize}

\begin{figure}[ht]
\centering
\includegraphics[width=1\columnwidth]{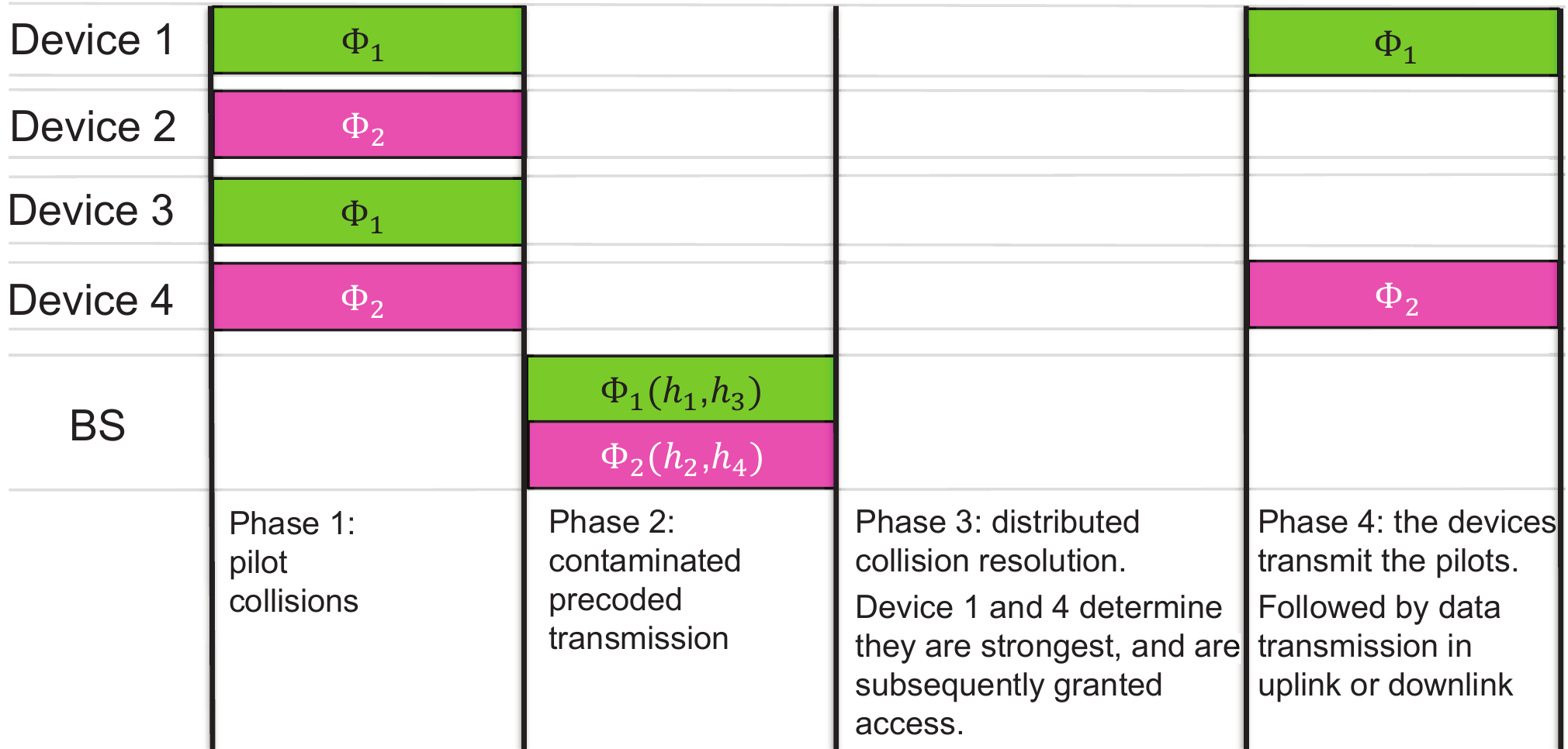}
\caption{Graphical representation of SUCRe protocol. 
Simplified example with 4 devices selecting a training sequence randomly from the set $\{\Phi_1,\Phi_2\}$. 
% \EDC{indicate in the figure: vertical line with phase 1, phase 2.. }. 
% In phase 1, 2 collisions ...
% In phase 2: better to write something "BS sends precoded ... with contamination". 
% Phase 3: Distributed collision resolution. Device 1 and device 2 are the strongest devices... 
% Phase 4: device 1 and device 2 are granted access to the pilots and proceed to collision free training. 
% You can indicate in the figure a data phase
}
\label{fig:sucre}
\end{figure} 

A graphical representation of how SUCRe works is illustrated in Figure~\ref{fig:sucre}.
The SUCRe protocol is able to resolve around 90\% of collisions, and is highly scalable in terms of number devices in the cell, since the collision resolution is decentralized.

\subsubsection{SUCRe extensions}

\begin{figure}[!htbp]
\centering
\includegraphics[width=.5\textwidth]{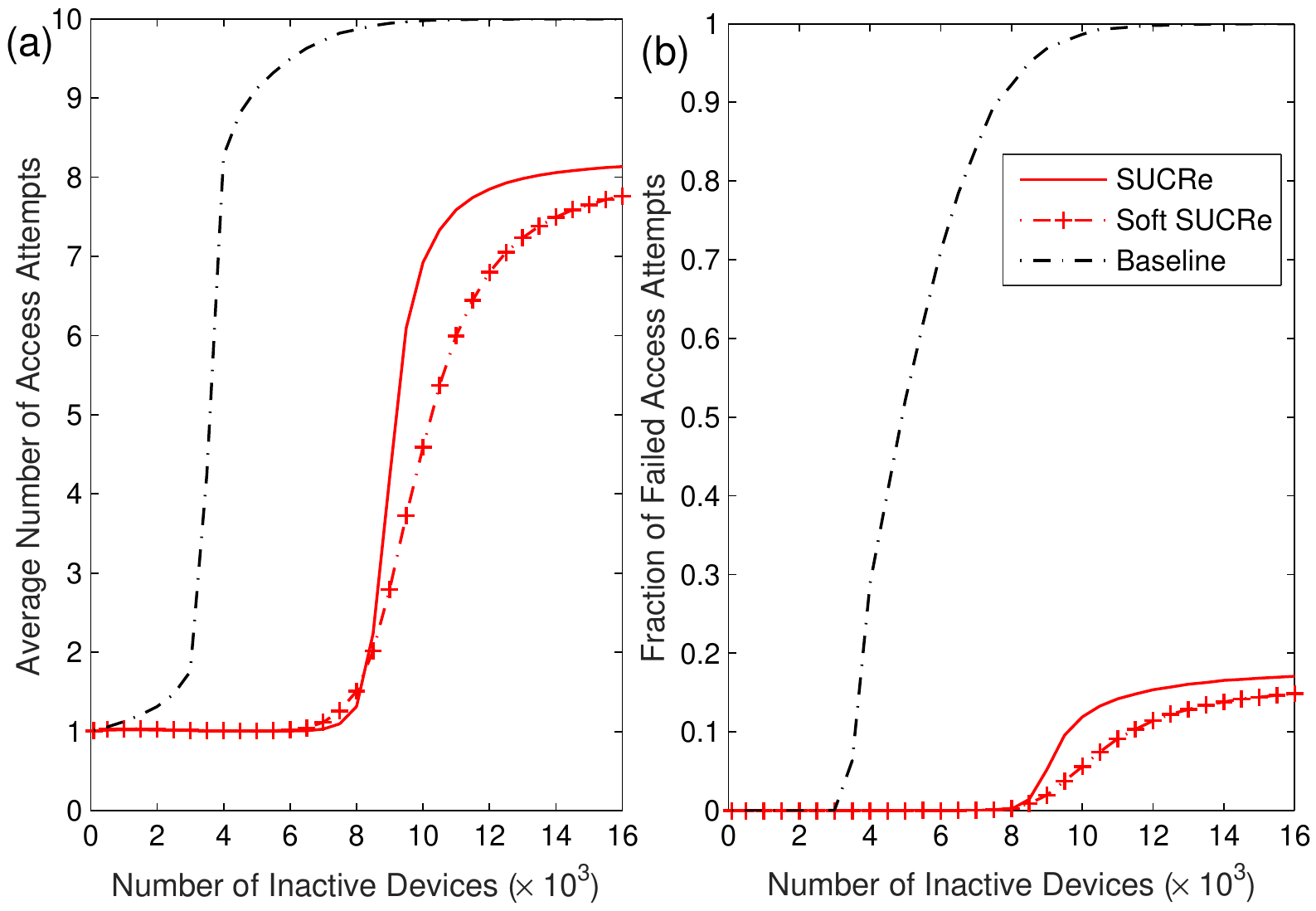}
\caption{RA performance in a crowded machine-type network without intercell interference: (a) Average number of RA attempts; (b) Probability of failed RA attempt as a function of the number of devices in a cell. Soft SUCRe approach computing the probability of being the strongest user.}
%\EDC{delete the curves that are not explained: interference (no need for that), baseline.}}
\label{fig:Perf_SSprob}
\end{figure}

The first extensions of the protocol, namely SUCR-IPA \cite{Han.2017} and SUCR-GBPA~\cite{Han.2017nov}, assign the pilots not used in {\it Phase 1} of  SUCRe to the devices that lost the collision resolution phase in {\it Phase 3}, at the cost of additional signaling.
In the classical SUCRe, the pilot collision is resolved with a hard decision for retransmission.
As the number of inactive devices $K_0$ becomes large, this hard decision is satisfied to a smaller extent as the number of collisions with a higher number of contending devices increases. 
For this reason, %the retransmission probability is proposed, as a \textit{soft decision retransmission rule}; such retransmission probability  is associated with the probability for a device to be the strongest contender and improves the performance of the SUCRe protocol in crowded scenarios \cite{Marinello2019}. 
the application of a {\it soft decision retransmission} rule by introducing the {\it retransmission probability} brings  improvement to the SUCRe in crowded scenarios \cite{Marinello2019}. 
Figure~\ref{fig:Perf_SSprob} depicts the average number of access attempts and the probability of failed RA attempts; an appreciable performance improvement can be obtained deploying the soft decision rule under the very crowded scenario, \emph{i.e.}, 
%$K_0 > \tau_p / P_a$
$K_0 > \mathcal{P}_{p} / P_a$
, where $P_a$ is the probability of activation.  
Besides, the reduction in the average number of access attempts of the soft-SUCRe of \cite{Marinello2019} in comparison with original SUCRe, indicates that 2.34 less RA attempts are required on average for 
%$K_0 = \tau_p/P_a = 10000$
$K_0 =  \mathcal{P}_{p}/P_a = 10000$
devices. Besides, the probability of failed access attempts in \mbox{Figure \ref{fig:Perf_SSprob}.b} can be reduced from $17.1\%$ to $14.9\%$ for 16000 devices. Baseline in \mbox{Figure \ref{fig:Perf_SSprob}} represents a conventional RA protocol with pilot collisions handled by retransmission in later blocks.
%It means that adopting the soft decision rule can provide successful access to $0.0213\times 16000 \times P_a = 0.3408$ additional devices on average if compared to the conventional SUCRe.}

%Figure~\ref{fig:Perf_St}.a depicts the probability of the SUCRe and  \colb{the {\it access class barring combined with a decentralized pilot power allocation} (ACBPC)  protocol \cite{JCM_TA_RD_PP_EdC2019}} of resolving pilot collisions as a function of the number of contending devices. It can be observed that for lower number of contending users, SUCRe is more effective, whereas if there is a higher number of contending devices, ACBPC is more effective.
%Moreover, the fairness of ACBPC %\EDC{defined?} is illustrated in Figure~\ref{fig:Perf_St}.b, where it can be noticed that the probability of a device winning a collision is less dependent on the proximity to the BS compared to SUCRe.
%It can also be noticed that soft-SUCRe incurs a slight performance degradation compared to classical SUCRe when the number of devices is not so high. 
%

\begin{figure}[htbp]
\centering
\includegraphics[width=.495\textwidth]{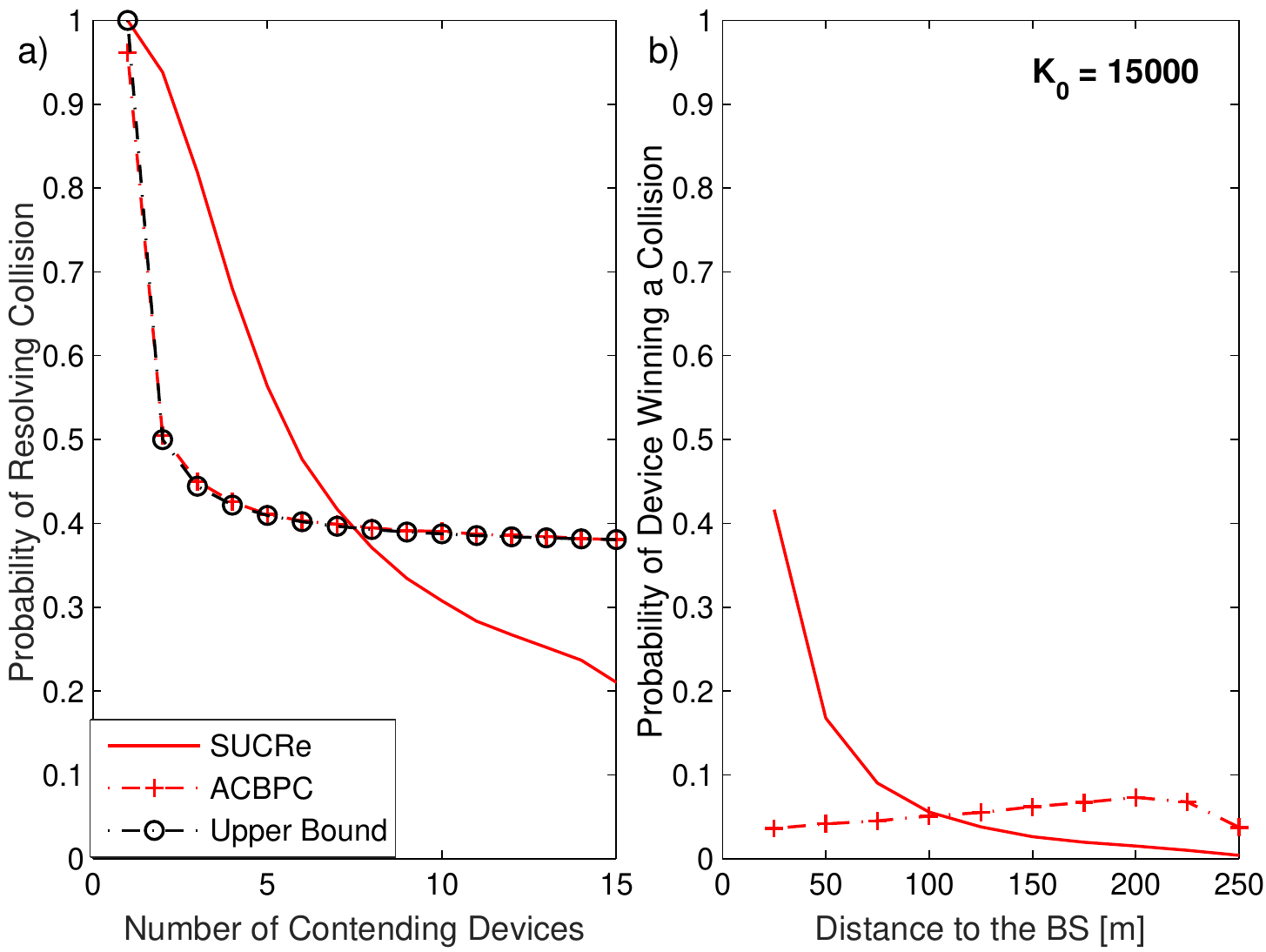}
\vspace{-5mm}
\caption{Grant-based ACBPC protocol: a) Probability of resolving collisions as a function of $|\mathcal{S}_t|$; b) Probability of a device winning a pilot collision as a function of its distances to the BS with $K_0 = 15000$.}
\label{fig:Perf_St}
\end{figure}

The original SUCRe protocol favors the access to the pilot sequences to devices with the lowest path loss, i.e. the highest signal. Therefore, collision resolution can be seen as unfair to the devices that are located far from the BS. SUCRe is extended in \cite{JCM_TA_RD_PP_EdC2019} providing an access class barring combined with a decentralized pilot power allocation (ACBPC) protocol that ensures a uniform RA performance for the devices within the cell, independently of their distances to the BS. 
Figure~\ref{fig:Perf_St}.a depicts the probability of the SUCRe and ACBPC %\EDC{defined?} 
of resolving pilot collisions as a function of the number of contending devices. It can be observed that for lower number of contending users, SUCRe is more effective, whereas if there is a higher number of contending devices, ACBPC is more effective.
Moreover, the fairness of ACBPC is illustrated in Figure~\ref{fig:Perf_St}.b, where it can be noticed that the probability of a device winning a collision is less dependent on the proximity to the BS compared to SUCRe.

%
% \begin{figure}[htbp]
% \centering
% \includegraphics[width=.495\textwidth]{figures/GraphPowC_ACB_ProbResColxSt_dist_2.eps}
% \vspace{-5mm}
% \caption{\colb{Granted-based ACBPC protocol}: a) Probability of resolving collisions as a function of $|\mathcal{S}_t|$; b) Probability of a device winning a pilot collision as a function of its distances to the BS with $K_0 = 15000$. \jcm{Please choose in the case of this Figure if the curves with intercell interference should or not be presented.}}
% \label{fig:Perf_St}
% \end{figure} 

\subsection{Grant-free random access protocols}
\label{sseq:grant-free}
Grant-free transmissions rely on transmitting both pilot sequences and data as part of the initial access attempt.
Grant-free protocols bring different advantages according to the type of pilot sequences considered, i.e. orthogonal and pseudo-orthogonal pilot sequences. In principle, the protocol is simplified as it does not resort to a dedicated collision resolution phase. 
However, in order to suitably decode the data from all the transmitting devices, improved transmission strategies or decoding strategies are needed. 

%\TA{In {\it grant-free} RA protocols, both metadata and data are sent in a single step given the opportunity to reduce the latency when compared to the {\it grant-based} access schemes. However, in {\it grant-free} RA protocols, device-activity detection is much more challenging since  orthogonal pilot sequences are not available for all machine-type devices due to a) the huge number of devices in mMTC crowded scenarios; b) finite wireless channel coherence time.}

%\subsubsection{With orthogonal pilot sequences}

Two protocols are described in this section. In both protocols, the transmission of the devices accessing the network is organized in multiple slots. 

\begin{figure}[t]
\centering
\includegraphics[width=1\columnwidth]{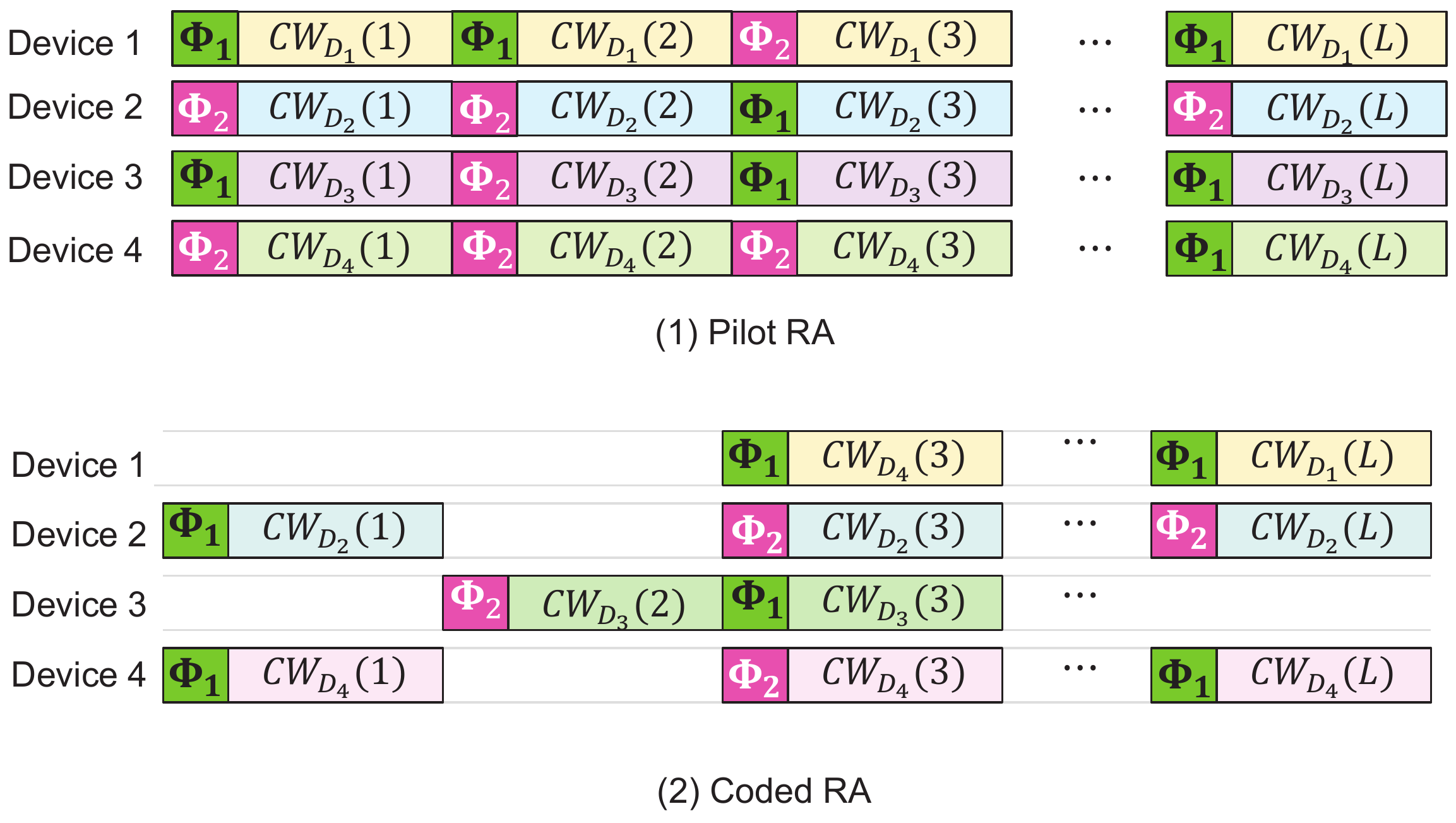}
\caption{Graphical representation of Pilot RA (1) and Coded RA (2). 
Simplified example with 4 devices, randomly selecting from a set of 2 pilots $\{\Phi_1, \Phi_2\}$. 
In (1) the codewords are divided into multiple parts, denoted $\{CW_{D_i}(1),\dots CW_{D_i}(L)\}$, transmitted in multiple time slots. In each time slot, a new pilot is chosen randomly.
In (2), the devices transmit in each slot with an activation probability, and choose the pilots based on a hopping pattern. The data in each time slot is the same, as in coded random access, such that SIC may be employed.}
\label{fig:RA}
\end{figure} 

\subsubsection{Pilot RA}
In the first protocol~\cite{Carvalho2017}, the codeword transmitted by a given device is divided into multiple parts. 
The device selects a new pilot sequence at random in each time slot followed by transmission of part of its codeword. In each time slot, the device might be involved in a pilot collision. 
The channel is assumed to be block fading, i.e. it takes an independent value during each transmission slot. 
It is estimated at each time slot and, due to pilot collision, channel estimation gets contaminated.
At reception, the BS employs maximum ratio combining (MRC) based on the contaminated channel estimate, creating interference in the data decoding phase. 
The main rationale in this approach is as follows. The pilot assignment hopping pattern creates an interference pattern in the data domain that changes at every time slot. With coding spreading across a large number of transmission time slots, the interference is averaged out. Hence, from an information-theoretic point of view, it is possible to define a reliable transmission rate. The approach is illustrated in Fig~\ref{fig:RA}1).
While the originally proposed protocol in \cite{Carvalho2017} relies on a priori (genie-aided)
information on the user activity, it is possible to blindly infer this information with high
reliability using sparsity-based signal processing techniques
\cite{BecirovicBL2019}.

\subsubsection{Coded RA}
In the second protocol~\cite{Sorensen2018}, the devices transmit with a certain probability of activation, with pilot hopping across time slots. The data is retransmitted in each time slot. 
This protocol is based on the principle of coded random access which can be described as follows.
By optimizing the activation probability, conditions for collision-free transmissions are created. Suppose that 
a collision free transmission happens at a given time slot, so that the data of the corresponding device is correctly decoded. Successive interference cancellation (SIC) is employed such that the contribution of the decoded packet is subtracted in the previous time slots thus possibly removing a transmission that was causing a collision. 
Applying the principle of coded random access to a massive MIMO system is not easy because it involves colliding transmissions in a joint channel estimation and data decoding problem. At each time slot, we estimate the channels using the pilot sequences that have been selected. MRC is applied based on the channel estimates that are possibly contaminated, thus producing interference in the data domain. The channel hardening properties of massive MIMO enable a formulation of a coded random access problem in the data domain based on the energy of the different channel involved. 

% \asb{also pilot contamination, half a column}
% \EDC{Coded random access: multi-slot transmission where the devices transmit with a certain probability, with pilot hopping at each slot and repetition of the data. Creates opportunities for collision free transmission. When a device transmission is collision free, it is decoded and its contribution is removed from previous time-slots creating additional opportunities of collision free transmissions in the past. \\
% Ergodic approach: in each time slot, pilot hopping and part of the data is transmitted. The interference pattern changes over time. With a long time horizon, the interference is averaged out and a reliable rate for transmission can be defined.}

%=========================
% \colr{(TA: under construction)}
% \vspace{5mm}
%=========================
\subsection{Grant-free access relying on compressed sensing}

%\EDC{
%o	With massive MIMO: in principle, it is possible to identify the training sequences present (by correlating), train the channels and build a multi-user receiver. \\
%o	Compressed sensing methods taking into consideration that the space occupied by the active devices is much smaller than the total spatial degrees of freedom available.}
%=========================
%\colr{(TA: under construction)}
%\vspace{2mm}

\begin{figure}[ht]
\centering
\includegraphics[width=0.75\columnwidth]{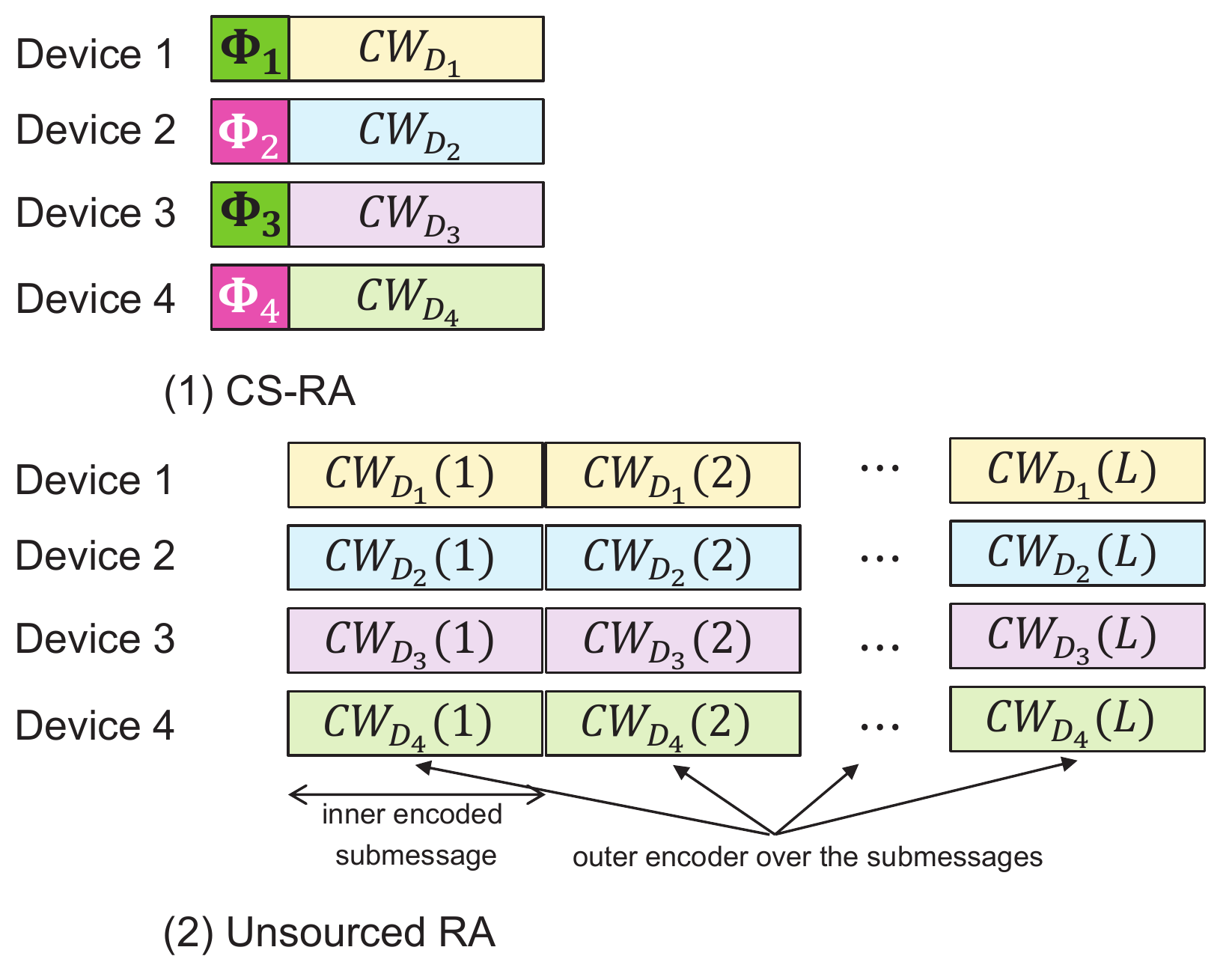}
\caption{Graphical representation of CS-RA and unsourced RA access.
In (1) CS-RA, pseudo-random pilot sequences are transmitted followed by data transmission.
In (2) UmRA, there are no pilots. Encoding is done using an inner encoder in each subslot and over multiple slots according to an outer encoder. Compressed sensing is employed on the inner low-dimensional codebook for lower complexity, whereas the outer code is used to reconstruct the overall message.
% Encoding: multi-slot transmission. compressed sensing on a low-dim codebook. outer code across subplot to reconstruct overall message. indicate this in the figure if possible: where compressed sensing is performed. where outer encoding has an effect. \EDC{rewrite}
}
\label{fig:RA2}
\end{figure}

Due to the sporadic nature of the MTC traffic, the device transmission patterns have been observed to become \textit{sparse}. 
Therefore, the  mMTC connectivity problem can be addressed from the perspective of compressed sensing (CS).
In this section, we consider a grant-free access where the data is transmitted in a single initial access attempt. {An important characteristic of the schemes discussed here is that a unique pilot sequence is assigned to each device, as illustrated in Figure~\ref{fig:RA2}a). Those sequences are pseudo-orthogonal as the number of orthogonal sequences is limited by the channel coherence time.
Therefore, obtaining the channel estimates by simply correlating the training received signal with each pilot is not an efficient approach, due to the non-orthogonality as well as due to the very large number of pilots.} We first describe schemes where training is performed jointly with data detection and then a scheme with non-coherent data detection. 
 
In the first type of access, compressed sensing techniques are employed  to facilitate reliable joint activity detection and accurate channel estimation.
%
%\EDC{General question: why don't we use a simple correlator with all possible pilot sequences to identify the devices and then perform channel estimation}
%\EDC{Questions: \\
%- in \cite{Chen18}, we have the assumptions "Random signature sequences with i.i.d. complex Gaussian distributed entries". Are the pseudo random sequences in the other references built in the same way? I think some of them are. Clarify.\\
%- I cannot  quite see why we make a distinction between  \cite{Chen18} and \cite{Liu2018, Liu2018b}. The differences I can see are:\\
%* \cite{Chen18} does not consider a data phase but \cite{Liu2018, Liu2018b} does. But this is trivial right? They have the same initial phase with activity detection + channel training. \\
%* \cite{Liu2018, Liu2018b} consider massive MIMO, ok. \\
%* \cite{Liu2018, Liu2018b} considered vector AMP and not \cite{Chen18}? \\
%* \cite{Liu2018, Liu2018b} makes channel estimation based on statistical channel information and not \cite{Chen18}. Which one? \\
%* what is the major distinction here?
%}
%
%In \cite{Chen18}, the {\it grant-free} mMTC connectivity problem is addressed from the perspective of compressed sensing. Random signature sequences with i.i.d. complex Gaussian distributed entries are assigned to the devices, and by exploiting the sparsity in the device activity pattern, the joint device detection and channel estimation problem is formulated as a compressed sensing single measurement vector (SMV) for single antenna BS, and as a multiple measurement vector (MMV) for MIMO BS. 
This compressed sensing problem is solved  via approximate message passing (AMP), with the aid of a novel MMSE denoiser function in \cite{Chen18} where a single antenna and a moderate number of antennas is considered at the BS.
%
%The authors propose to resolve the formulated problems via approximate message passing (AMP) algorithm, with the aid of a novel MMSE denoiser function, and also derive analytical performance expressions for the investigated system. However, the scenario adopted in \cite{Chen18} considers only moderate number of antennas at the BS.
%
An extension where the BS is equipped with a very large number of antennas is proposed in \cite{Liu2018, Liu2018b} based on  AMP.
%(\EDC{is it correct to assume that \cite{Chen18} does not use vector AMP.} \jcm{\cite{Chen18} refers to the technique employed in the multiple-antenna case as AMP with vector denoiser, but comparing eq.(8) and (9) from \cite{Chen18} with eq.(13) and (14) from \cite{Liu2018}, one can see it is the same vector AMP technique.})
%\jcm{As an extension of \cite{Chen18}, a compressed sensing solution for grant-free access with massive MIMO is} proposed in \cite{Liu2018, Liu2018b}. A two-phase access scheme is considered, in which device activity detection
%\EDC{How is it done without pilot sequences which serve to identify the devices}
%and channel estimation are performed jointly using non-orthogonal pilot sequences %\jcm{with i.i.d. complex Gaussian distributed entries} in the first phase and data is transmitted in the second phase. 
%{Although the total number of devices is very large, the number of active devices is kept of the same order as the number of BS antennas in order to achieve appreciable data rates.} 
The performance of the vector AMP algorithm with MMSE denoiser based on statistical channel information in the asymptotic massive MIMO regime is analyzed in \cite{Liu2018}. %analyzes the performance of the vector AMP algorithm with MMSE denoiser  based on statistical channel information in the asymptotic massive MIMO regime, 
It is shown that it can achieve perfect device activity detection performance as the number of BS antennas grows indefinitely, with both misdetection and false alarm probability going to zero.
%\EDC{in view of the next sentence, I guess this regime assumed that the number of active users in much smaller than the number of antennas? If so, mention it clearly.}
However, in \cite{Liu2018b}, it is demonstrated that the overall achievable rate attained by this scheme in such scenario 
%(\EDC{in non-asymptotic conditions?} \jcm{in asymptotic conditions; in non asymptotic, both channel estimation errors and device activity detection limit achievable rates.}) \jcm{in asymptotic conditions} 
is limited by the increased channel estimation errors, which result from the deployment of non-orthogonal pilots in order to accommodate a larger number of devices.

The massive MIMO grant-free access is further investigated in \cite{GF_mMTC_maMIMO_CS}. The employed non-orthogonal pilot sequences are generated by sampling an i.i.d. complex Bernoulli distribution.
%, \emph{i.e.}, each entry follows $(\pm 1 \pm j)/\sqrt{2 \tau_p}$, with $\tau_p$ representing the pilots length. 
Using such pilots and the AMP algorithm for device detection along with channel estimation, authors show that using the acquired channel estimates for coherent data transmission is suboptimal. A better approach is to use the AMP algorithm only for device activity detection, and then using conventional MMSE channel estimation for active devices. Authors also demonstrate the suboptimality of employing identical pilot transmission power for all devices, while simple power control strategies, based on long-term fading coefficient inversion, allow substantial performance improvements. Most important, they propose a novel non-coherent data transmission technique, 
%\EDC{not clear why this is non-coherent. You need to explain more.}
which encodes $r$ information bits to be transmitted onto $2^r$ possible pilots per device. A modified AMP algorithm is also proposed aiming at exploiting the sparsity incurred by the  non-coherent transmission scheme. {Note that the proposed scheme is said to be non-coherent because there is no need to obtain explicit channel estimates. The entire process consists of only detecting which pilots are active, while the ID of the sender as well as the information bits are implicit on the index of the pilot sent.} Performance comparison revealed that the proposed non-coherent transmission significantly outperforms coherent transmission scheme, being thus a promising approach for future mMTC networks.

%========================
\subsection{Unsourced Massive Random Access {(UmRA)}}
%=========================
%\EDC{what is the basic principle of unsourced random access (citing polyanskiy)? Devices have the same transmission protocol: what does it mean?
%The principle: the devices cannot be identified based on their transmitted signals. We are interested in decoding the messages while identification of the transmitting devices is not possible.}
%\EDC{what are the assumptions about CSI? in the original work? In a paper by Petar (ISIT), the assumption is that each device knows its own CSI and inverts the channel. In Caire's work: non-coherent detection? So each device transmits a message from the same codebook. No training is sent, I guess. At the receiver, the message are received with different fading coefficients at each antenna and for each device}
% \EDC{what is the principle of decoding in that case in simple terms. I guess it is different from the case described in section C.}

%\EDC{to be inserted sooner. In mMTC random access, a very large number of devices operate in an uncoordinated fashion. At any time, only a small subset of the total are active, i.e., with some payload to be sent. In this setup, conventional random access solutions assume activation detection of the transmitting devices and the fact that the probability of successful joint decoding goes asymptotically to one with increasing blocklength. However, in the context of mMTC the devices have small data payloads. Even though the subset of devices that are active simultaneously can be rather small, the large total number of devices results in finite blocklength (FBL) effects.}

In many mMTC applications targeting, for example, alarm and large-scale sensor reporting, only the content of the message is relevant, and not the identity of the sender. This is in stark contrast to conventional way one thinks about wireless communications. The rationale is that the actual data is important rather than who is sending it. 
A novel information-theoretic treatment for this scenario was provided by Y. Polyanskiy in \cite{Polyanskiy2017}. 
In this model, a small number of active devices transmit using the same codebook, which precludes the identification of devices. Later on, the approach by Polyanskiy has been termed  \textit{unsourced random access} \cite{Vem2017}. 

The probability of misdetection and false alarm is also different from the classical definition that declares error if any one of the messages is decoded incorrectly. 
A per-device error probability is defined as the average fraction of transmitted messages that is not in the list of decoded messages at the receiver. Likewise, a per-user probability of false-alarm is defined as  the average fraction in the list of decoded messages  that have not been transmitted. 

In the context of mMTC the devices have small data payloads. Even though the subset of devices that are active simultaneously can be rather small, the large total number of devices results in finite blocklength (FBL) effects.
The unsourced, uncoordinated nature of the problem and the FBL effects have implications on the design of practical low-complexity coding schemes. In \cite{Polyanskiy2017}, Polianskiy provided bounds on the performance of finite-length codes. In \cite{Kowshik2019}, a practical low-complexity iterative LDPC scheme for a fading channel is proposed. 

A compressed sensing approach is proposed in \cite{Fengler2019a} to identify the transmitted messages. The received signal can be described as a linear mixture of codewords from the same codebook. 
Sparsity  arises since  the  size of the {codebook} is very large (exponential with the number of bits per submessage, allowing a one-to-one correspondence between submessages and codewords) and the device activity is sparse.  Therefore, decoding of the linear mixture can be performed using compressed sensing methods. Due of the large size of the codebook, such methods appear infeasible. Instead, an alternative solution leading to a lower complexity is proposed in \cite{Fengler2019a} based on a two-step encoding with an inner common encoder and an outer common encoder. 
The transmission slot is partitioned into subslots. In each subslot, a device transmits part of the codeword, a submessage, that is encoded using an inner encoder, as illustrated in Figure~\ref{fig:RA2}. The transmission of the whole message across the subslots is ruled by an outer encoding.  
In each subslot, the  inner decoder identifies which submessages have been transmitted. Because the device activity is sparse, the decoding corresponds to a sparse problem that is solved based  on a compressed sensing method with a low complexity as the size of the codebook is reduced compared to the encoding previously described.
The outer decoder reconstructs the list of transmitted messages based on the list of sub-messages 
%\PP{define a sub-message}. 

Unsourced Massive Random Access {(UmRA)} {\cite{Fengler2019}}  is the extension of the previous encoding scheme to massive MIMO. The increased number of measurements provided by the large number of receive antennas is exploited to improve the inner decoding at each subslot.  
Unlike \cite{Fengler2019a}, the issue raised by channel estimation in massive MIMO is explicitly considered by employing non-coherent activity detection (sub-message detection). 
This approach is non-Bayesian and relies on the covariance matrix of the vectorial received signal. It does not assume a priori knowledge about large-scale pathloss coefficients of the devices or about the activity patterns. This is an advantage over Bayesian approaches {\cite{Fengler2019a}} based on message passing where such assumptions are made and can be seen as unrealistic.

\section{Massive MIMO and network slicing}
\label{seq:network_slicing}

\begin{figure}[t]
    \centering
    \includegraphics[width=1\linewidth]{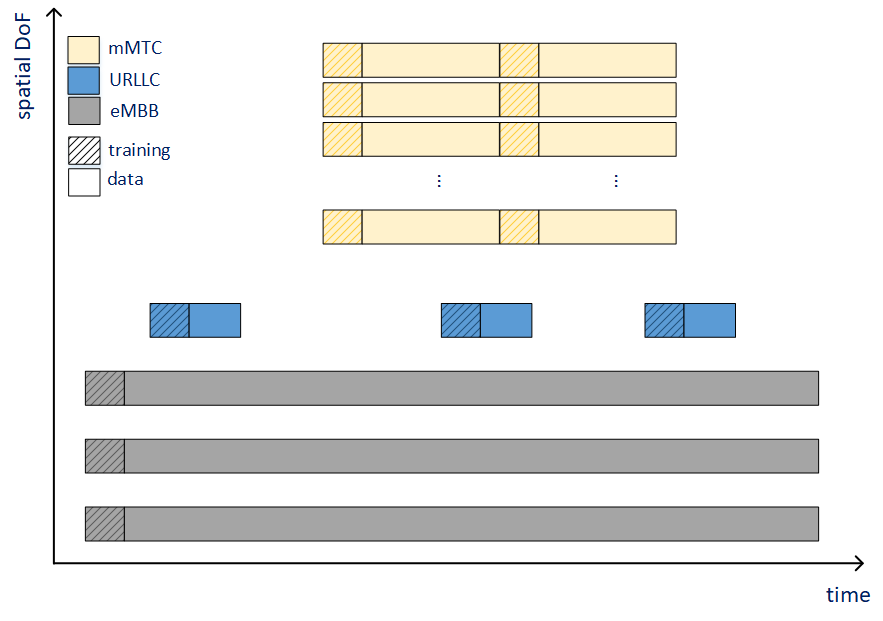}
    \caption{Example of orthogonal network slicing of the wireless resources in a time-space grid for supporting eMBB, URLLC and mMTC.}
    \label{fig:networkslicing_mMIMO}
\end{figure}
As introduced in Section \ref{sec:3gpp}, network slicing is a key feature of 5G for the support of heterogeneous services. In the RAN, the conventional approach to slice is to allocate orthogonal radio resources to eMBB, with very long payloads; mMTC, characterized by the large amount of devices and the need of a random access; and URLLC, with small packets and low latency requirements \cite{Popovski2018}. The multiplexing in space of such services with very different characteristics and requirements brings major challenges. Figure \ref{fig:networkslicing_mMIMO} shows an example of the slicing of the three services in the time-spatial grid.  

\subsection{Training}
While massive MIMO is conventionally used to separate intra-service data traffic, it can also be an efficient tool for network slicing and isolate the services based on multi-antenna processing. 
 Nevertheless, separating the services with multiple antennas relies on a prior training phase where the CSI is acquired, which significantly complicates the design and the co-existence of services. In the following, we discuss the case of TDD, being much more mature than massive MIMO FDD solutions.

%
%\subsubsection{Orthogonal and pseudo-random pilots}
In TDD, the CSI is usually estimated using uplink training with orthogonal pilot sequences. eMBB devices share a pool of orthogonal pilot sequences, $\mathcal{P}_{B}$. A different pool of orthogonal pilot sequences is required for URLLC, $\mathcal{P}_{U}$. 
Since the URLLC packet has to be sent as early as possible, so does the URLLC training which will not be synchronized with eMBB in general as there is no time to wait until the next eMBB training slot. Therefore, there is no need to impose mutual orthogonality between $\mathcal{P}_{B}$ and $\mathcal{P}_{U}$. However, URLLC training should be performed such as to minimize the interference from other services. 

For mMTC, we can have the case of orthogonal pilot sequences and the case of pseudo-random pilot sequences. 
With grant-free access, training and data are sent together in the single access attempt. For the BS, this means doing a joint training and collision resolution, which requires another pool of  pilot sequences. Those sequences could be orthogonal, but, on some implementations, the orthogonality of the set can be compromised by the large amount of devices. A solution is to rely on pseudo-random sequences: one unique sequence is assigned per user and serve to identify each user. The disadvantage of pseudo-random sequences is that it creates interference in the training phase. 
For the orthogonal pilot sequence, a grant-based access can be used (see section~\ref{sseq:grant-based}) %such as Aloha
, where the training can be performed once the device gets the right to use a pilot sequence. 
%the channel. In that case, the pilot sequences may be created by a pseudo-random generator that uses the unique device identifier as seed. \asb{A second possibility is} if the BS has a list of devices and their unique identifiers, then the pilot sequences associated with each device are known at the BS. 
We refer to the pool of mMTC sequences as $\mathcal{P}_{M}$, either orthogonal or pseudo-orthogonal.  %\asb{this seems a bit contradictory to table \ref{table:RAprotocols}, where grant-free can be with both orthogonal and pseudo-orth.}. 
If $\mathcal{P}_{B}$ and $\mathcal{P}_{M}$ are mutually orthogonal, then the training can be aligned.  determines the alignment of the training; otherwise, they will be separated in time. 
%\PP{Not sure what you want to say here. Tee orthogonality is determined by the alignment or vice versa?}

\subsection{Training alignment}

When considering the three services simultaneously, the channel acquisition is seriously affected by the inter-service interference. The training can be broadly organized to be aligned or not aligned among services, being \mbox{Figure \ref{fig:networkslicing_mMIMO}} an example of the latter. 

%If full training is required, there are two strategies: to have periodic training, such that the system is always ready when the data arrives; or to preempt on-going long transmissions (e.g., eMBB traffic) and do the URLLC training upon request.   

%For URLLC, the only option is not aligned, and puncturing of on-going eMBB and mMTC might be required to ensure an interference-free training. This has an extra implication for mMTC, whose training shall necessarily be allocated before the first URLLC request arrives, to ensure that the data streams can be separated. 

For multiplexing mMTC and eMBB, \cite{Senel2018} compares the option of having not aligned, aligned or time-multiplexed training phases. The latter means that mMTC are not allowed to transmit during the training period of eMBB, which reduces the eMBB pilot length. After the training of eMBB, mMTC transmit their pilot sequences followed by data transmission, which is proven to give performance gains.  

Furthermore, the latency requirements of URLLC puts strict constraints to the training. Things get more challenging depending on the transmission direction, particularly when the URLLC request is in the downlink whereas the training has to be done in the uplink. If full training is required, one solution is to perform periodic training, such that the system is always ready when the data arrives.

\subsection{Data phase}
In the data phase, the key challenge is in the coexistence of services, particularly when they have opposite directions. A relevant case is a long eMBB transmission going on e.g., in the DL. If the URLLC request is also in the DL, puncturing-like solutions like the ones under consideration in 5G can be adopted \cite{Popovski2018}. The key assumption is that the reliability of URLLC is two or more orders of magnitude higher than eMBB, such that eMBB requirements can be fulfilled even with the loss due to the puncturings. If the URLLC request is instead in the UL, the eMBB transmission shall be preempted to allocate the URLLC device. 

%Degrees of freedom available. Can we accommodate mMTC in space-slicing 
%mMTC can take a huge amount of resources in the slicing, and coexistence with eMBB is probably feasible only for a limited number of simultaneous active devices (e.g., no more than XX active devices for YY antennas).

\subsection{FDD Massive MIMO with heterogeneous services}
 Most of the massive MIMO research has focused on TDD and channel-reciprocity. For FDD bands, massive MIMO can be exploited by means of a predetermined grid of beams with devices reporting their preferred beams. It has been analytically shown that with isotropic scattering (independent Rayleigh fading) pilots-based TDD outperforms the FDD option \cite{Flordelis2018}. However, the industry keeps a high interest in massive MIMO solutions for FDD, motivated by spectrum regulations. Certainly, using FDD bands would simplify the network slicing operation, eliminating the  challenges associated to the opposite transmission directions.

% Some perspective on FDD: network slicing advocates for massive MIMO FDD. We have to make it work with eMBB.

%%%%%%%%%%%%%%%%%%%%%
\section{Conclusions and Directions}
\label{seq:concl}
%Massive array processing provides considerable gains, suitable for machine-type communication. 

Since its inception about a decade ago, Massive MIMO has evolved from a ``wild academic idea'' into commercial reality and a key technology component for sub-6 GHz wireless access in 5G. What will come next?  In all likelihood, refinement of the basic Massive MIMO will continue, for example by learning and exploiting statistical information about the propagation environment, and spatial correlation in particular, and by learning and exploiting user behaviors and traffic patterns.  This pertains especially to random-access protocols, where huge improvements can be achieved by exploiting this type of side information.  Also, incorporating context awareness into the protocols, for example in terms of specific packet deadlines that are dependent on the eventual use case, is bound to provide significant boosts.  Ultimately, the freshness of information may be quantified through the information-theoretic concept ``age of information'' \cite{kosta2017age}.

More importantly, however, the next major leap in wireless access technology is likely to involve so-called cell-free Massive MIMO technology \cite{ngo2017cell,nayebi2017precoding}.  The cell-free architecture fundamentally departs from the cellular (Massive MIMO) paradigm, as follows. A number of access points are spread out geographically and are not generally equipped with large antenna arrays; they may (preferably) have a single or possibly a small number of antennas. Theoretically, all access points may participate in the service of every terminal through phase-coherent beamforming, though in practice only the closest access points will effectively contribute. There is no concept of cells; all resources, including pilots, are reused universally. Access points are connected via backhaul to central processing units. The distinction between cell-free architectures and the (already commercialized) concept of ``small cells'' \cite{andrews2014will} is important: Small cells do not cooperate coherently, whereas the access points in a cell-free architecture do. Random access in cell-free Massive MIMO will be a challenge. In particular, while this paper has discussed a large array of different techniques that have shown to be greatly promising in \emph{cellular} Massive MIMO, it is not clear how or to what extent they generalize to cell-free Massive MIMO.  Conceptually, one can think of cellular Massive MIMO as a special case of cell-free Massive MIMO where the path loss from a given terminal to every base station antenna is the same - whereas in contrast, in cell-free Massive MIMO it is different (and differs by many orders of magnitude).  Channel hardening, for example, a fundamental phenomenon in cellular Massive MIMO upon which some algorithms discussed here rely to some extent, does not hold to an equal degree in cell-free Massive MIMO. The development of grant-free and efficient random access procedures in cell-free Massive MIMO remains a grand challenge that will be utterly important for beyond-5G systems.

New visions about wireless access have recently emerged. Those visions bank on promising technological progress that will enable an easy deployment of thin electromagnetic panels of very large physical dimensions. Those panels would be active, i.e they can transmit and receive electromagnetic signals. Such visions have appeared in recent years under different names such as extremely large aperture massive MIMO \cite{Amiri2018GC}, extra-large scale massive MIMO \cite{EDC2019} or large intelligent surfaces \cite{EdforsLIS2018}. Those panels can be an integral part of new large infrastructures such as a stadium or inside an airport. They are envisioned as thin, flexible stripes \cite{Interdonato2019} or surfaces that can be fixed on walls and easily powered up. 

Because of its physical size and proximity to the wireless devices, it is well understood that such electromagnetic panels can bring a significant performance boost in the area they cover, but they also have the potential to play a prominent role in achieving low latency and reliable communications. Let us take the example a large factory where it is essential to communicate at very low error probability. Imagine covering the walls, ceilings or floors with electromagnetic panels. This type of deployment is quite different from conventional massive MIMO that usually cover wider areas and do not enclose the communicating devices in such a way. Those electromagnetic panels offer the unique capability to capture a complete 3D ultra-high resolution snapshot of the environment. The purpose can be to communicate with the aid of side information brought by tracking the structure of the channel or detect sudden anomalies calling for changing in communication patterns.

%%%%%%%%%%%%%%%%%%%%%

%\bibliographystyle{IEEEtran}
%\bibliography{refs_mMTC.bib}
\input{main.bbl}

\end{document}